\documentclass[12pt,preprint]{aastex}

\newcommand\ha{H${\alpha}$~}

\slugcomment{The Astrophysical Journal, submitted}

\shorttitle{Dust in Magellanic Cloud Planetary Nebulae}
\shortauthors{Stanghellini et~al.}
  
\begin{document}
  
\title{{\it Spitzer} Infrared Spectrograph Observations of Magellanic Cloud Planetary Nebulae: 
the nature of dust in low metallicity circumstellar ejecta\altaffilmark{1}}

\author{Letizia Stanghellini}
\affil{National Optical Astronomy Observatory, 950 N. Cherry Av.,
Tucson, AZ 85719}
\email{lstanghellini@noao.edu}

\author{Pedro Garc\'\i a-Lario}
\affil{Herschel Science Centre, European Space Astronomy Centre, Research and Scientific Support Department of ESA, 
Villafranca del Castillo, P.O. Box 50727. E-28080 Madrid, Spain}
\email{Pedro.Garcia-Lario@sciops.esa.int}

\author{D.~Anibal ~Garc\'\i a-Hern\'andez}
\affil{The W. J. 
McDonald Observatory. The University of Texas at Austin, 1 University Station C1402, Austin, TX 78712}
\email{agarcia@astro.as.utexas.edu}

\author{Jose V.~Perea-Calder\'on}
\affil{European Space Astronomy Centre, INSA S.~A., P.O. Box 50727. E-28080 Madrid, Spain}
\email{Jose.Perea@sciops.esa.int}

\author{James E. Davies}
\affil{National Optical Astronomy Observatory, 950 N. Cherry Av.,
Tucson, AZ 85719}
\email{jdavies@noao.edu}

\author{Arturo Manchado}
\affil{Instituto de Astrof\'isica de Canarias, v\'{\i}a L\'actea s/n, La Laguna, E-38200 Tenerife, 
Spain; affiliated to CSIC, Spain}
\email{amt@iac.es}

\author{Eva Villaver}
\affil{Space Telescope Science Institute, 3700 San Martin Drive, Baltimore MD 21218; 
affiliated to the Hubble Space Telescope Department of ESA}
\email{villaver@stsci.edu}

\author{Richard A.~Shaw}
\affil{National Optical Astronomy Observatory, 950 N. Cherry Av.,
Tucson, AZ 85719}
\email{shaw@noao.edu}

\altaffiltext{1}{based on observations made with the {\it Spitzer Space Telescope},
which is operated by the Jet Propulsion Laboratory, California Institute of Technology, under a contract with NASA.}

\begin{abstract}
We present 5--40 $\mu$m spectroscopy of 41 planetary nebulae (PNe) in the 
Magellanic Clouds, observed with the Infrared Spectrograph on board the \textit{Spitzer 
Space Telescope}. The spectra show the presence of a combination 
of nebular emission lines and solid-state features from dust, superimposed on the thermal IR continuum. 

By analyzing the 25 LMC and 16 SMC PNe in our sample we found that
the IR spectra of 14 LMC and 4 SMC PNe
are dominated by nebular emission lines, while the other spectra 
show solid-state features. We observed that
the solid-state features 
are compatible with carbon-rich dust grains (SiC, polycyclic aromatic hydrocarbons (PAHs), etc.) in most cases, 
except in three PNe showing oxygen-rich dust features. The frequency of carbonaceous dust features is generally higher in
LMC than in SMC PNe.

The spectral analysis allowed the correlations of the dust characteristics with the
gas composition and morphology, and the properties of the central stars. We found that:
1) all PNe with carbonaceous dust features have C/O$>$1, none of these being bipolar or otherwise
highly asymmetric;
2) all PNe with oxygen-rich dust features have C/O$<$1, with probable high mass progenitors
if derived from single-star evolution (these PNe are either bipolar or highly asymmetric); 
3) the dust temperature tracks the nebular and stellar evolution; and
4) the dust production efficiency depends on metallicity, with low metallicity environments not favoring 
dust production.

\end{abstract}

\keywords{Planetary nebulae; Magellanic Clouds; Dust; solid state features; morphology}

\section{Introduction}

Planetary nebulae (PNe) are the envelopes ejected toward the end of the evolution 
of low- and intermediate-mass stars (1--8 M$_{\odot}$), at the tip of the thermally pulsing 
asymptotic giant branch (TP-AGB) phase. The gas and dust ejected in this phase 
contain elements that had been produced by nucleosynthesis and then carried to 
the stellar surface by convective dredge-up processes. 
The slow wind that 
eventually produces the PN, also called the superwind, starts with the AGB thermal 
pulses and carries most of the envelope mass during the last few thermal pulses 
(Vassiliadis \& Wood 1993). The composition of the ejecta is tightly correlated to the
initial mass and metallicity of the evolving star: higher masses in this range will have undergone
hot bottom burning (HBB) nucleosynthesis, which depletes carbon and 
increases nitrogen in PNe (e.g., Marigo 2001).
The exact mass-loss mechanism and superwind onset are unclear to date, and, most importantly, 
their dependence on the mass and metallicity of the stellar progenitor is unconstrained.

Dust is formed on the cool surface of the AGB star, and radiation pressure on the dust grains 
may be a dominant factor in the superwind onset. Furthermore, the nature of the
dust grains affects dust opacity, thus the role of dust grains in the superwind efficiency 
must be related to the chemistry of the nebula. Interestingly, Willson (2000) showed that
mass loss might also occur in the absence of dust. 

A PN is born when the circumstellar matter ejected during the superwind phase is 
ionized by the UV flux from the central star. 
Dust in PNe survives the sputtering due to the UV radiation field for a time period
that depends on the nature of the dust grains, rather than on the expansion rate and the 
luminosity of the central star (e.g., Kaeufl et al.\ 1993). 
Dust in Galactic PNe has been observed extensively. It has been shown that about 40\% of the emergent bolometric 
flux in Galactic PNe occurs between 5 and 60 $\mu$m (Zhang \& Kwok 1991) in the form of a blackbody-like 
continuum spectrum due to dust scattering (Cohen \& Barlow 1974). Dust in Galactic PNe has been 
characterized by its cool (100--200 K) temperature, producing a spectrum peaking at 25--60 $\mu$m.
The physical extent of the dust coincides with the gas 
emission in most nebulae and is even more extended in others (Volk 2003).

Observations of AGB stars and PNe in the Galaxy and the Large (LMC) and Small (SMC) Magellanic Clouds, and
other Local Group galaxies provide increasing evidence that metallicity may play a
role in the AGB and beyond (i.~e., Garc{\'{\i}}a-Hern{\'a}ndez et al. 2007). 
On average, the LMC and SMC metallicities are respectively $\sim$50\% and 25\% that of the Galaxy
(Caputo et al.\ 1999).
The absence of heavily obscured AGB stars in the Magellanic Clouds (Groenewegen et al.\ 2000; Trams et al.\ 1999), 
in contrast with
their Galactic counterparts, suggests that, on average, lower metallicity environments
are less favorable to dust production. 
The ratio of carbon-rich to oxygen-rich AGB stars in Local Group galaxies decreases with increasing metallicity
(Cioni \& Habing 2003; Cioni et al. 2003; Schultheis et al. 2004). 
AGB stars in low-metallicity galaxies would take a long time to 
lose their envelopes if the mass-loss efficiency was proportional to dust content.
Their lifetime as AGB stars could be very long, increasing the chance of the
nuclear-processed material to surface in their lifetime. In this way, a higher 
concentration of nuclear-reaction products is explained.
In contrast, high-metal stars would have powerful dust-driven mass-loss rates.
As a consequence, their chemistry will be less affected.
Stanghellini et al. (2003) analyzed all Magellanic Cloud PNe that have been previously observed with the
{\it HST} cameras and found that the frequency of asymmetric (e.~g., bipolar, quadrupolar) PNe
in the SMC is only 60$\%$ that
of the LMC, where the metallicity is, on average, higher. 
One may infer that bipolar
morphology is less favorable in the low-metal environment of the SMC than in the 
LMC and the Galaxy.
Aspherical PNe, especially the bipolar ones, are typically more massive
than their spherical counterparts, indicating a more efficient mass-loss mechanism (Peimbert \& Serrano 1980). 
Furthermore, the mere bipolar shape indicates the presence of a gas-dust torus around the central star,
constraining the outflow into bipolar shape. It is no coincidence that the internal extinction of
bipolar PNe is higher than that of spherical PNe. It occurred to us that
the different morphologies may 
form with completely different mass-loss mechanisms, and that the low-metallicity, low-dust
environments may produce only spherical PNe.

From the above described observing results, it is obvious that metallicity affects the 
properties of PNe, so that it becomes clear that the dust properties in PNe 
need to be studied in a variety of
metallicity environments, and there is a good reason to observe dust features in 
relation to the absolute physical characteristics of the nebulae and stars. 
These are the scientific strands that motivated us to perform the IRS observations on the
Magellanic Cloud PNe presented in this paper.
While Galactic PNe have been studied with IRAS and ISO 
(Zhang \& Kwok 1991, Garc{\'{\i}}a-Lario \& Perea Calder{\' o}n 2003), 
most Magellanic Cloud PNe were beyond the reach of these IR telescopes.
Magellanic Cloud PNe
have known distances, thus lowering the number of free parameters in the 
comparison between the data and the models. 

This paper presents low resolution IRS {\it Spitzer} spectra for 25 LMC and 16 SMC 
PNe. In \S2 we present the observations relative to this data-set and the data 
analysis, including the spectral analysis and a description of 
individual nebulae. Section 3 presents the study of the correlations between 
the dusty nature of the PNe and their gas morphology and evolution, $\S$4 
contains the discussion of our findings, 
and $\S$5
summarizes our conclusions and suggests future endeavors in this field.

\section{Observations and data analysis}

\subsection{Sample selection}

Our goal is to
study the infrared spectral properties of PNe in the Magellanic Clouds, and to relate them
to morphology and other nebular  properties, and to the nature of their central stars,
in order to gain insight on their formation and evolution.
In order to meet our science goals we have selected LMC and SMC PNe
that satisfy the following selection: 
(1) All targets have been previously observed with the {\it HST} cameras (Shaw et al. 2001, 2006; Stanghellini et al. 1999, 2002), 
thus morphological (and, in many cases, stellar) information is available.
(2) All targets are smaller than 2\arcsec\ in apparent size, 
to facilitate accurate peak-up observations. This requirement excludes only a very small fraction
of Magellanic Cloud PNe that have been observed with {\it HST}. These restrictions do not 
exclude evolved PNe significantly: at the distance of the Magellanic Clouds, 2\arcsec\ implies a PN diameter between 0.5 pc (for the LMC) and 0.6 pc (for the SMC). Thus, for typical PN expansion velocities our sample  includes PNe up to $\sim$9000 yr after ejection (see models by Villaver et al. 2002), which is a substantial fraction of their visible lifetime. 
(3) The abundances of the elements that are related to AGB evolution and population (He, N, O, Ar, S) 
are available in the literature. 
(4) Carbon abundances have been measured from their UV spectra (Stanghellini et al. 2005, Leisy \& Dennefeld 1996).

After filtering the Magellanic Cloud PNe sample through the above selection, we also excluded those targets already in the Reserved Observation Catalog (ROC) through the {\it Spitzer} GTO or other programs. We shall study those targets in the future through the {\it Spitzer} Data Archive. In Figure 1 we show the histogram of the 
[\ion{O}{3}] $\lambda$5007 flux distribution of all Magellanic Cloud PNe that satisfy selection rules (1) and (2)
above. These include targets observed by us (black histogram), and those in the ROC (gray histogram). 
The sample presented in this paper is most representative 
of the [\ion{O}{3}]-bright Magellanic Cloud PNe, where the [\ion{O}{3}] $\lambda$5007 brightness is a good proxy for the 
optical PN brightness for all but the most unevolved PNe (Stanghellini et al. 2002). 
However, there is increasing evidence for a significant population of very low excitation (VLE) PNe in both the LMC (Reid \& Parker 2006b) and the SMC (Jacoby \& De Marco 2002), where [\ion{O}{3}] is very weak and the 
F([\ion{N}{2}])/F(\ha) ratio can be very strong. While it is not clear 
that a sample selection based on [\ion{O}{3}] brightness inherently selects 
against a particular population of PNe, VLE nebulae may nevertheless be 
under-represented in our sample. 
Additional mid-IR spectra of VLE nebulae will be needed to understand the 
applicability of the results presented here to this class of PNe. In Table 1 we
list the selected targets, with their angular radii and morphologies derived
from the {\it HST} images (Shaw et al. 2001, 2006; Stanghellini et al. 1999,
2002).

\subsection{Spitzer Observations}

The spectral data were acquired with the Infrared Spectrograph (IRS, Houck et al.
2004) on board the {\it Spitzer Space Telescope} (Werner et al. 2004) between 2005 July 14 and
November 14 as part of a General Observer program ($\#$20443).The observing log is given in Table 2, 
where we list the target name, the coordinates from the {\it HST}
images, the observing date, the dataset ID number, and the observing campaign. The 5--40 $\mu$m range 
spectra were acquired by using the Spitzer 5.2--8.7 $\mu$m Short Low 2nd order (SL2), the
7.4--14.5 $\mu$m Short Low
1st order (SL1), the 14.0--21.3 $\mu$m Long Low 2nd order (LL2), and the 19.5--38.0
$\mu$m Long Low 1st order (LL1) IRS modes. Each of the observations consists of 3 cycles of 14 sec
and 30 sec duration each for the SL and LL segments respectively.
We also performed peak-up on a nearby star to achieve
accurate (0.4$''$) pointing. We obtained spectra of 41 PNe. The target SMP~LMC~30, originally on our target list,
was not pointed accurately enough for spectral extraction.

The data were first processed by the {\it Spitzer Science Center} (SSC) using the
standard data reduction pipeline version 12.4 (except for one source, SMP~LMC~100,
which was processed by the pipeline version 13, which was suitable for this dataset). This process includes the 
 linearization correction, the subtraction of darks, and cosmic ray
removal. The images were also corrected for stray light, and a flat-field correction was applied 
to account for pixel-to-pixel response variations. 
Through the IRS data reduction pipeline we obtained three Basic Calibrated Data (BCD) files
for each slit nod position. These three cycles were then combined by the pipeline
using a signal-weighted average to produce the co-added 2-D images. 

The pipeline-processed 2-D images were further cleaned by interpolating over the
flagged data and rogue pixels, which depend on the observation campaign. 
Sky background was removed by using the image degeneracy relative to
the two nod positions. Spectral extraction, and wavelength and flux
calibration were performed with the {\it Spitzer IRS Custom Extractor} (SPICE), with point source
aperture. The resulting 1-D spectra were cleaned for bad pixels, spurious jumps and
glitches, smoothed and merged into one final spectrum per module for each source
using the IRS Spectroscopy Modeling Analysis and Reduction Tool 
(SMART\footnote{SMART was developed by the
IRS Team at Cornell University and is available through the {\it Spitzer Science Center}
at Caltech.}). The overlap regions
between orders in the four modules were masked by hand to produce a smooth,
artifact-free spectrum in the 5.2--38 $\mu$m range. The
third {\it bonus} order contributed by the SL and LL modules were also masked, so in
the end, each PN has two spectra, one for each nod position.

\subsection{Spectral analysis}

The infrared emission from the target PNe is composed of one or more of the following 
components: narrow emission lines from collisionally excited atomic species (nebular 
lines), thermal continuum emission, and  
emission from dust grains. Most of our IRS spectra show continuum emission at different 
brightness levels, and nebular emission lines and solid state features superimposed 
on the continuum. We analyzed each of these components in turn. 

We identified and measured the nebular lines in the PN spectra by using the line-fitting 
routine in SMART. A single-order fit was made to the baseline for each nebular line, 
and the line was then fitted using a single Gaussian. The measured line strengths and 
widths will be presented in a future paper, together with a detailed nebular analysis. 
We found a spurious feature in the spectrum, nominally at 19 $\mu$m, whose 
detection was possible due to its slight offset between nodding positions. In some 
cases it falls very close to the [\ion{S}{3}] line at 18.7~$\mu$m, and in these cases the 
feature was removed before the measurement of the [\ion{S}{3}] line.
From an inventory of the nebular emission lines in the IRS wavelength range we 
defined four major excitation classes. We designate the PNe to be low excitation if 
only the [\ion{S}{3}] lines are visible, intermediate excitation when the spectra shows 
[\ion{Ar}{3}], [\ion{S}{4}], and [\ion{Ne}{3}] nebular emission, high excitation if we also 
detect [\ion{O}{4}], and very high excitation if we also see the [\ion{Ne}{4}] or 
[\ion{Ne}{5}] emission. Several PNe show both low and high excitation lines, likely 
due to different excitation zones in the nebulae. 

In order to study the emission from the dust, a single blackbody was fitted to the 
merged cleaned continuum using data from both nod positions. Nebular lines and 
solid-state features were masked during the fitting procedure to isolate the continuum. 
Data with $\lambda >38 \mu$m in the LL1 module were often masked as well due to 
the increased noise at these wavelengths. Figures 2 through 5 show the best
fit Black-body curves to the data, which are adopted as the continuum 
fits, and the continuum-subtracted spectra in four PNe representing a variety of IRS spectral features . 
The top panels show, for each nod position, the original spectra and the Black-body fit (indicated by a thicker 
line); the bottom panel shows the continuum-subtracted spectra and includes the 
emission line identifications. 
Solid-state features were measured using an IDL routine developed by us which 
allows for varying the start and end-points of the feature to be fitted. Solid-state fits 
were performed on the continuum-subtracted spectra.

All the IRS spectra show evidence of a thermal continuum with superimposed 
nebular emission lines. In addition, the presence of solid-state emission features
are superimposed on some of the spectra. 
We sort the PNe into three different IRS spectral types: F, CRD, and ORD types. 
We define as F (for featureless) those PNe that show no evidence of solid-state features. 
The lack of features is 
due either to the actual absence of grain emission, such 
as may be expected in evolved PNe, or to the dust features possibly being below the 
detection limit. In this class, we include the objects that show possible but not certain 
dust emission features. 
CRD (for carbon-rich dust) PNe are those whose spectra show carbon-rich dust features. This class includes multiple 
types of dust grains. Among CRD we find spectra showing the very broad 11 $\mu$m 
emission of silicon carbide, 
other spectra showing broad peaks at 6-9, 11-15, and 15-21 $\mu$m generally attributed to very small
carbonaceous grains, possibly formed by large PAH clusters.
PNe with evidence of the classic PAH emission features at 3.3, 6.2, 7.7, 8.6, and/or 
11.3~$\mu$m are classified as CRD. Some CRD PNe show both broad and 
narrow carbon-rich dust features, superimposed, as if the ionized PAH features were 
beginning to form out of the clusters of small carbonaceous grains as evolution 
proceeds. 
ORD (for oxygen-rich dust) PNe are those whose spectra show oxygen-rich dust emission features, such as crystalline 
silicates. 

In Figures 6 through 8 we show all the observed IRS spectra. Figures 6a, 6b, and 6c include all F
PNe (listed alphabetically, the LMC PNe first). 
Figures 7a, 7b, and 7c show the CRD
PNe. In Figures 7a through 7c we order the PNe by the strengths of the carbonaceous dust features, as they generally 
are thought to evolve 
in PNe. The sequence shows the evolution of certain features, but it is not an evolutionary sequence
since PNe belong to different mass and/or metallicity sequences.
Figure 8 shows the small sample of ORD PNe, where PNe have been ordered 
according to the strength of the degree of processing of their silicate features. All spectra show the two nod positions
averaged. In Figures 7 and 8 these averages were normalized with respect
to the flux at 14 $\mu$m, a feature-free region of all spectra.

In Table 3 we summarize the results from the analysis of the IRS spectra. In column (2) 
we give the dust temperature of the continuum, fitted as a black body; column (3) 
gives the infrared flux derived from the integration of the continuum dust spectra 
between 5 and 40~$\mu$m; column (4) gives the IR excess; 
column (5) gives their dust type; and 
column (6) gives the excitation class of the PN, based on the observed emission 
lines. 
From Montecarlo simulation based on several fits for the continua we infer that the fitting errors
are below $\sim$5$\%$ both in temperatures and IR fluxes.
To obtain the ratios in column (4) we have used the Balmer line fluxes of these 
PNe from Stanghellini et al. (2002, 2003), and Shaw et al. (2006).

Below we list the general characteristics of the observed PNe. The dust type,
based 
on the infrared spectrum for each target, is included.

\noindent{\bf SMP~LMC~4} (Fig.~6a, F): With a low continuum level, the spectrum of this PN is regular
and does not show solid-state features with the exception of very faint emission at 
11.3 $\mu$m, which could be of interstellar nature, or the remnant of evaporated dust. The spectrum is almost featureless, 
but its classification based on solid-state features is uncertain.
The nebular emission lines are characteristic of a very high excitation spectrum.

\noindent{\bf SMP~LMC~9} (Fig.~7b, CRD): This spectrum shows faint emission 
at 11.3 $\mu$m, which is likely to be from PAHs. A host of nebular emission lines show a very high excitation
nebula by the presence of [Ne V] at 24.2 $\mu$m.

\noindent{\bf SMP~LMC~10} (Fig.~6a, F): A featureless spectrum apart from the nebular emission lines, including
high excitation ones.

\noindent{\bf SMP~LMC~16} (Fig.~6a, F): The line at about 9.8 $\mu$m could be molecular 
hydrogen at
$\lambda$9.66 $\mu$m, which is in contrast with the very high excitation nebular lines. 

\noindent{\bf SMP~LMC~18} (Fig.~6a, F): Faint and featureless, with the exception of a possible
broad dust feature at 30$\mu$m.
The spectrum shows intermediate excitation nebular lines. Its classification based on dust features is uncertain due to 
a possible very low detection of the 15--20$\mu$m broad feature.

\noindent{\bf SMP~LMC~19} (Fig.~7b, CRD): The feature at 11.3 $\mu$m is likely to be narrow PAH emission.
 The emission spectrum indicates very high excitation.

\noindent{\bf SMP~LMC~20} (Fig.~6a, F): As in the spectrum of SMP~LMC~16, we are possibly detecting
molecular hydrogen at 9.66 $\mu$m. Again, as in SMP~LMC~16, the nebular lines denote a very high excitation PN, thus the
molecular line is rather improbable.

\noindent{\bf SMP~LMC~21} (Fig.~8, ORD): This PN has a unique spectrum that shows strong crystalline
silicate emission. The nebular emission line spectrum is also 
interesting, showing a combination of low (i.e., [S III])
excitation nebular emission lines with very high excitation
                 ones (i.e., [Ne VI]).

\noindent{\bf SMP~LMC~25} (Fig.~7a, CRD): This spectrum shows carbon-rich dust features, which are likely 
due to PAH clusters, or very small carbonaceous grains. We observe peaks at 6.2, 7.7, and 8.5 $\mu$m, all
likely to be proto-PAH features, and a huge SiC emission at 11.3 $\mu$m. 
 This is an intermediate excitation nebula.

\noindent{\bf SMP~LMC~27} (Fig.~6a, F): This is an intermediate excitation nebula with no dust features, with the possible exception
of the 30$\mu$m broad feature as in SMP~LMC~18.

\noindent{\bf SMP~LMC~34} (Fig.~6b, F): An almost featureless spectrum, except for the nebular emission lines
typical of intermediate excitation, and the solid-state feature at 30$\mu$m, as in
SMP~LMC~18. It is hard to classify it based on solid-state features.

\noindent{\bf SMP~LMC~45} (Fig.~6b, F): The nebular spectrum shows low to high excitation features, 
while the solid-state features are missing with the possible exception of the 30$\mu$m feature.

\noindent{\bf SMP~LMC~46} (Fig.~7b, CRD): The features at 7.9 $\mu$m and 11.3 $\mu$m could be PAHs or PAH-related, but very faint.
The nebula has intermediate to high excitation.

\noindent{\bf SMP~LMC~48} (Fig.~7a, CRD): The spectrum shows several complex solid state features. The broad feature
at 6--9 $\mu$m, possibly due to small grain clusters, is superimposed on narrow features at 6.2, 7.7, and 8.6
$\mu$m, characteristic of PAHs. Other features are also seen at 15--20 $\mu$m and 30$\mu$m. 
The PN is of intermediate excitation from the nebular
lines.
 
\noindent{\bf SMP~LMC~66} (Fig.~6b, F): In this nebula it is most apparent that the feature at
29.6 $\mu$m is an artifact, since it only appears in one of the nodding
positions. This is a high excitation nebula with no solid state features
in the spectrum.

\noindent{\bf SMP~LMC~71} (Fig.~7c, CRD): This is a very high excitation spectrum with a
solid state feature
at 6.2 to 8.6 $\mu$m, possibly due to small grain clusters, superimposed on narrow features at 6.2, 7.7, and 8.6
$\mu$m, characteristic of PAHs. Other features are also seen at 15--20 $\mu$m and 30$\mu$m. 

\noindent{\bf SMP~LMC~72} (Fig.~6b, F): This is a very high excitation spectrum without solid-state features.

\noindent{\bf SMP~LMC~79} (Fig.~7c, CRD): Very interesting PAH features are present at 6.2, 7.7, 8.6, 11.3, and 13.5 $\mu$m, 
superimposed on broad features. Spectral lines are characteristic of a high
excitation PN.

\noindent{\bf SMP~LMC~80} (Fig.~6b, F): Dust featureless spectrum with intermediate excitation emission lines.
The thermal continuum is undetected in this PN.

\noindent{\bf SMP~LMC~81} (Fig.~8, ORD): The spectrum presents evident silicate features (SiO) at 9--12.5 $\mu$m.
The broad emission between 16 and 30 $\mu$m is probably due to dust continuum. 
The nebular emission spectrum shows intermediate excitation lines.

\noindent{\bf SMP~LMC~95} (Fig.~6b, F): This nebula shows a high excitation spectrum. Although almost featureless, the 
solid-state spectrum is
characteristic of an evolved PN, with a possible faint PAH feature showing at 11.3 $\mu$m.

\noindent{\bf SMP~LMC~97} (Fig.~6c, F): This is a very high excitation PN without solid state features, with the possible exception
of the 30$\mu$m broad feature.

\noindent{\bf SMP~LMC~99} (Fig.~7c, CRD): This object shows a very high excitation spectrum. The solid-state features are similar to those
of SMP~LMC~71, with narrow PAH emission features superimposed on the broad dust features, possibly due to clusters of
small grains. The 30$\mu$m broad feature is evident in the spectrum.

\noindent{\bf SMP~LMC~100} (Fig.~7c, CRD): A very high excitation PN, this spectrum shows broad PAH signature and the 11--15 $\mu$m bump
typical of unevolved carbon-rich dust.

\noindent{\bf SMP~LMC~102} (Fig.~6c, F): There is a possible PAH detection in this otherwise featureless
spectrum.

\noindent{\bf SMP~SMC~2} (Fig.~7b, CRD): The spectrum of this very high excitation PN
shows very likely PAH emission at 11.3 $\mu$m. 

\noindent{\bf SMP~SMC~5} (Fig.~7c, CRD): Several PAH features are seen in this PN, superimposed on broad carbon-rich dust
emission peaking at 8, 11.3 $\mu$m. The nebular excitation is very high. 

\noindent{\bf SMP~SMC~8} (Fig.~6c, F): The spectrum is featureless from the viewpoint of solid-state 
compounds. The nebular excitation is intermediate.

\noindent{\bf SMP~SMC~9} (Fig.~6c, F): The spectrum shows no features and no emission lines, and the two nod
position do not have the same continuum level. 

\noindent{\bf SMP~SMC~13} (Fig.~7a, CRD): This intermediate excitation spectrum shows broad carbon-rich features with peaks
around 8 and 12 $\mu$m, typical of carbon-rich unevolved dust grains, and the broad feature at 30 $\mu$m.

\noindent{\bf SMP~SMC~14} (Fig.~7c, CRD): A high excitation PN, shows narrow PAH features 
with peaks at 6.2, 7.7, 8.6, and 11.3, characteristic of PAH emission, superimposed on broad features.

\noindent{\bf SMP~SMC~15} (Fig.~7a, CRD): The [\ion{S}{4}] $\lambda$10.5 emission appears on top of 
the broad solid state feature. 
This nebula is an extremely good example of the emission expected from carbon-rich dust grains before 
they
become PAHs. Note that the 15--21 $\mu$m feature may be the precursor of the PAH feature that would
be found at 17.0--17.7 $\mu$m. In addition, the broad 30 $\mu$m feature is also clearly seen. 
This PN is an intermediate excitation one.

\noindent{\bf SMP~SMC~16} (Fig.~7b, CRD): 
Broad carbon-rich dust features appear with peaks at 6--9 and 11-15  $\mu$m, typical of unprocessed clusters 
of small carbonaceous grains and/or large PAH clusters. This is a low excitation PN. 
 
\noindent{\bf SMP~SMC~17} (Fig.~7c, CRD): Similar to SMP~SMC~16, but with PAH narrow peaks appearing 
at 6.2, 7.7, 8.7, and 11.3 $\mu$m.
This is an intermediate excitation PN. 

\noindent{\bf SMP~SMC~18} (Fig.~7a, CRD): This is another very good example of the emission 
expected in a carbon-rich source with dust grains where PAHs have not yet been formed, showing broad SiC emission. 
Note the plateau features at 6--9 $\mu$m, 11-15 $\mu$m, and 15--21 $\mu$m, as well as the 
extremely broad 30 $\mu$m feature. The excitation for this nebula is intermediate.

\noindent{\bf SMP~SMC~19} (Fig.~7c, CRD): Some carbon-rich dust features, both narrow PAH peaks and broad, unevolved ones, 
are present, together with very high excitation lines.

\noindent{\bf SMP~SMC~20} (Fig.~7a, CRD): This source is similar to SMP~SMC~15; with huge SiC features, it is probably
the least evolved source among the PNe showing carbon-rich features.

\noindent{\bf SMP~SMC~23} (Fig.~6c, F): This object has a featureless spectrum, apart from a couple of 
intermediate excitation emission lines. 

\noindent{\bf SMP~SMC~25} (Fig.~8, ORD): Unique source in the SMC showing oxygen-rich dust silicates peaking 
shortward of 11 $\mu$m. The solid state features are similar to those of 
SMP~LMC~81, but with some crystalline silicates features arising over 
the continuum longward of 20 $\mu$m. This is a very high excitation 
nebular spectrum, high
dust and effective temperature, and a very high central star mass PN.

\noindent{\bf SMP~SMC~26} (Fig.~6c, F): The faint feature at 11.3 $\mu$m could be a PAH. The nebular spectrum shows very
high excitation lines.
     
\noindent{\bf SMP~SMC~27} (Fig.~7b, CRD): Intermediate excitation PN with carbon-rich dust emission of the broad type.

\section{Correlations between IR spectral characteristics and other nebular and stellar properties}

We studied the interrelations between the properties that we derived from the IRS spectra and the 
properties derived from other studies. We have specifically selected a target sample of PNe whose 
morphology, gaseous carbon abundance, and central stars have been studied
through {\it HST} imaging and spectroscopy. The sizes of the PNe have been determined from the 
angular size measured from {\it HST} images (see Table 1) and the distances to the Magellanic Clouds. In this paper we 
adopt d$_{\rm LMC}$ =50.6 kpc (Freedman et al. 2001; Mould et al. 2000)
and d$_{\rm SMC}$ = 58.3 kpc (Westerlund 1997), used also to get the IR 
luminosity in solar units from the fluxes of Table~3.

The PNe in our sample represent an assortment of initial masses and metallicities; they show a variety
of morphological types and also represent a variety of 
evolutionary stages. Care is needed in disentangling the evolutionary effects on the physical quantities 
plotted and their intrinsic values, related to mass and metallicity.

Figure 9 shows the measured infrared luminosity versus carbon abundance (here the abundances are in terms of log X/H +12,
where X is the element considered)
of all PNe of our sample that have a carbon determination either from Leisy \& Dennefeld (1996)
or from Stanghellini et al. (2005). We estimated that the infrared luminosity errors are of the order of 5$\%$;
typical carbon abundance uncertainties are $\sim5$ to 10$\%$. This implies that errorbars in Figure 9
would be smaller or comparable to the sizes of the symbols used for the data points. 
The two parts of Figure 9 display PNe on the same plane, 
where dust type and morphology are identified by symbol type in the left and right panels respectively.
The loci of the different dust type PNe are well separated in this
plot: the PNe with featureless spectra (F), are typically at low to intermediate luminosity and present 
a spread in carbon
abundances. The PNe with CRD spectra are on the upper right of the
plot. PNe with oxygen-rich dust spectra, or ORD, are on the upper left of the plot. From the figure 
we infer a
strong relation between the gas and dust composition: CRD IRS spectra are only
found in carbon-rich PNe.

By comparing the left and right panels of Figure 9
we note that no bipolar, quadrupolar, or point-symmetric PNe (squares)
in the figure correspond to PNe with carbonaceous dust. The carbon
 gas abundance, dust type, and morphology are thus very clearly interrelated.
Since gaseous carbon is depleted in massive PN progenitors via the HBB reactions, 
in Figure 9 the progenitor mass of PNe decreases from left to right. 
 PNe with high mass progenitors
 in the figure are ORD PNe if they are IR-bright;
 otherwise they are featureless. It is possible that some of the F PNe on the lower left of the plot
 are indeed evolved versions of the ORD PNe on the upper left. 
 It is worth noting that the SMC PNe in the plot are restricted to the upper right part, with log C/H +12$>$8.4 and
 log L$_{\rm IR}$/L$_{\odot}>$1.9.
 
 More insight comes from Figure 10, where we plot the carbon vs. oxygen
abundances (oxygen abundance uncertainties are similar or smaller than
those for carbon, thus
errorbars in Figure 10 are smaller than symbol size). Once again, the left and right panels identify dust type and morphology.
The left panel shows that PNe with PAHs and other 
 carbonaceous dust features all have C/O$>$1, while C/O$<$1 for all
 ORD PN, and there are no exceptions to these correlations. 
The right panel of Figure 10
 shows that all asymmetric PNe have low carbon abundance, and either oxygen-rich dust or
no dust features. It is clear that PNe whose progenitors have gone through the HBB
do not develop carbonaceous dust. We found that the average N/O for CRD PNe is $\sim$0.3, while
it is $\sim$2 for ORD PNe, a clear sign of HBB activity. 

In Table 4 we summarize the group properties of the different dust types, separating the columns for the LMC and the SMC 
PNe. In this table, abundances (Stanghellini 2006) are reported linearly, in terms of hydrogen. 
In column 1 we give the diagnostics,
and columns (2) through (7) give respectively their values (and standard deviation, in row 2 through 6) for the 
samples. Note that in a few cases we do not have data to fill the tables, and in those cases where we did not report 
the deviation 
 we only have one datum for the diagnostics.
It is worth mentioning that ORD PNe, those whose progenitors are likely more massive and which have gone through 
HBB nucleosynthesis, are
not necessarily those with the higher oxygen abundances. This is a phenomenon that is especially marked in 
the Magellanic Cloud PNe, where the enhanced efficiency of the 
ON cycle at low metallicity depletes both carbon and oxygen in favor of nitrogen (Stanghellini 2006). This is also why 
the
oxygen abundances are a poor indicator of PN metallicity for PNe whose progenitors have gone through the 
HBB.

In Figure 11 we show the dust temperature vs. physical radius of the 
nebulae (both quantities have uncertainties smaller than 0.02 dex).
The straight line shows the best fit using all plotted data, and suggests a 
power-law decline of T$_{\rm dust} \propto$~R$^{-0.25}$. 
Once again, dust types are identified in the left panel, morphology in the right panel. 
The physical radii of these PNe are photometric radii (e.g., Shaw et al.\ 2001), 
and correspond to the geometrical radius of the ionized gas for all except the bipolar
PNe with large lobes. Planetary nebulae with larger radii are generally more evolved 
than those with small radii, thus it is not surprising that the radii are inversely 
proportional to some power of the dust temperature. Lenzuni et al. (1989) showed 
a similar plot for Galactic PNe that were observed with the IRAS satellite and 
inferred that T$_{\rm dust} \propto~$R$^{-0.15}$. 
We confirm that the grain efficiency of PNe (within the same dust type) decreases with 
time, probably due to sputtering of material off the grain surface. It is also
interesting to note that our much higher precision in the radius determination (because in 
Magellanic Cloud PNe, the distances are known and accurate angular
sizes are available from the HST imagery) and the dust temperature (since temperature is derived from 
a spectral fit, and not by interpolation of a few photometric data points, as for the IRAS 
data for Galactic PNe) does not significantly decrease the scatter of the relation compared 
to the Galactic plot of Lenzuni et al. (1989); this suggests that the scatter is real and depends 
on the grain properties.

In Figure 12 we show the tight relation between the dust temperature and the IR luminosity. PNe evolve
from high to low dust temperatures, and from high to low IR luminosity, i.e.,
from the upper right to the lower left of the plots. All ORD PNe are at intermediate 
evolutionary stages. Uncertainties of the data points are smaller than 0.02 dex in
both axes. 
At very low IR luminosity and temperatures, where the most evolved PNe are found,
we only find F PNe. 

The electron density of the gas is another parameter that decreases as the nebula evolves. This is studied 
in Figure 13, where the dust temperature and the electron density show a tight correlation.
 It is worth noting that the CRD PNe follow a tighter correlation than F and ORD PNe.
In is also important to mention that dust temperature does not correlate with optical extinction. Typical errors
in the logarithmic density scale are smaller than $\sim$0.1 dex, while the dust
temperature uncertainties are negligible in this scale.

While the anti-correlation of dust temperature with nebular radius indicates a general evolutionary trend, it is only by
studying nebulae and their central stars together that one can disentangle the evolutionary effects from the
effect of the mass of the PN progenitors and the final core mass of the central stars. In the HR diagram of Figure 14
we can
characterize the evolution of post-AGB stars by mass bins, thus singling out the effects of evolution from those
due to dynamical evolution and nebular acceleration.

In Figure 14 we show the loci of the central stars, of those PNe in our sample whose stellar physical
parameters are available, on the log T$_{\rm eff}$--logL/L$_{\odot}$ plane
(individual effective temperature errors
are negligible on the log scale (see Villaver et al. 2003, 2004, 2006), 
while luminosity uncertainties are typically
of the order of 0.05 dex).
The plot shows the evolutionary tracks 
of hydrogen-burning post-AGB stars with masses of 0.57, 0.6, 0.7, 0.,75, and 0.9 M$_{\odot}$, taken from
the models of Stanghellini \& Renzini (2000). The 
symbols refer to the dust type 
(left panel) and morphological (right panel). We have plotted only stars whose luminosity has been measured directly, 
from their magnitudes, and have not
included model-dependent luminosity determination (Villaver et al. 2003, 2004, 2006). We only used
those stars whose effective temperatures have been evaluated via He II Zanstra analysis. From the figure we infer that
the F and CRD PNe can be found around stars with a
range of 
evolutionary stages and core masses, while the only ORD PNe whose location on the HR diagram
is known correspond to a high mass evolutionary track.

In Figure 15 we show the sequence of CRD PNe with nebular size. All PNe whose spectra have been depicted in Figure 7
(a through b) have been assigned a sequence number, 1 to 20, as ordered in the Figure. The sequence starts with PNe 
featuring huge SiC
(in order of strength), then goes to those PNe with broad and narrow carbonaceous features superimposed, then 
to the PAH feature PNe, and finally to those PNe with strong PAH emission. 
The symbols in Figure 15 refer to nebular morphology with the usual coding. We infer that the nebular
expansion is related to the carbonaceous dust sequence, linking nebular evolution and dust
evolution in particular for round and elliptical PNe. 
The scatter represents the PN spread in progenitor mass and metallicity.

\section{Discussion}

In our analysis we found that 56\% of the Magellanic Cloud PNe studied present solid state features, 
superimposed on the thermal dust continuum. These solid state features are, in 87\% of cases, the 
signature of carbonaceous dust, with only a few PNe showing oxygen-rich dust emission instead.
The rest of the nebulae show only nebular emission lines on a thermal continuum, in some cases 
very low.
By contrast, such a low continuum and absence of features has
not been observed in Galactic PNe (Garc{\'{\i}}a-Lario et al., in preparation).
It is remarkable that in the low-metallicity environments of the Magellanic Clouds
the solid state features seem to be rarer than in Galactic PNe.
This simple observation shows that the metallicity of the population has an enormous impact 
on the dust that forms on the AGB, as we knew already from carbon star counts in the Local Group Galaxies. What is
new here is that the nature of the dust in PNe is different for different environments. Mass-loss, if caused by pressure 
on the dust grains, must vary accordingly. The selection effects are different for Galactic and Magellanic Cloud PNe,
and a quantitative comparison of the Magellanic Cloud and Galactic samples is beyond the scope of this paper, but it
is important to note that the selection effects that affect the LMC and SMC samples are comparable,
thus the two samples readily comparable.

The ratio of CRD versus ORD PNe is 11 in the SMC, and 4.5 in the LMC, compared to $\sim$1 in the Galaxy
(Garc{\'{\i}}a-Lario et al., in preparation).
Selection effects limit the quantitative impact of the comparison between 
Magellanic Cloud PNe and the Galactic results.
The evolution of carbonaceous dust proceeds from the amorphous state, to crystalline, to PAH; the grains
evolve from small clusters of molecules to ionized PAH, as the central star becomes hotter (Peeters et al. 2004;
Moutou et al. 2000). The grains become bigger
but eventually evaporate under the action of the UV radiation field; thus, at late evolutionary stages
we would not observe much dust in PNe. 
Not only is the dust chemistry different in different environments, but the evolutionary stage of the 
dust grains is also statistically different. For example, the SiC compounds are found only in SMC PNe and not
in the LMC (or their Galactic) counterparts. In our sample
there are three SMC PNe (SMP~SMC~15, SMP~SMC~18, and SMP~SMC~20)
with broad SiC emission, but none in the LMC. Given that the LMC sample is larger than the SMC sample,
this is remarkable and of course needs to be studied with larger samples in both galaxies.
A few other SMC and LMC PNe have broad carbonaceous dust emission, such
as SMP~SMC~13, SMP~LMC~25, and SMP~LMC~48. 
Overall, features of amorphous dust grains are more
common in the SMC than in the LMC PNe and are very uncommon in the Galaxy. 
If these structures really are the early evolutionary stages
of PAHs, then we have found that the PN evolutionary rate depends strongly on population metallicity.

We found that there are strong correlations between the dust type, the carbon abundance, and the
IR integrated luminosity. CRD and ORD PNe are generally IR-bright, and at the 
opposite extremes of the carbon abundance range. All CRD PNe have C/O$>$1, and 
all ORD PNe have C/O$<$1. 
Most of the F PNe are low IR-luminous objects, suggesting that 
some of them might be an evolved version of the other type of PNe. 
All CRD PNe are either round or elliptical in shape, some with a bipolar core, but none of them show
 extended lobes or other marked asymmetries. The opposite is true for ORD PNe.
 
The dust temperature correlates well with other physical parameters that change with evolution,
 (e.g., the nebular radius, the IR-luminosity, and the electron density). CRD PNe define a tight sequence
 on the log T$_{d}$--log R plane (where R is the photometric radii of the PNe) similar to those found in Galactic PNe. 
 The CRD PNe whose spectra
 shows amorphous carbonaceous dust grains, such as SiC, have small physical radii,
 high electron density, and high IR-luminosity, in agreement with their being the unevolved version of PNe 
 with PAH emission. Our empirical evolutionary sequence of carbonaceous dust features correlates well with the
 physical dimension of the nebulae, and their dust temperatures. Other correlations of IRS spectral type with
 excitation classes, and a modeling approach to these data will be thoroughly discussed in a future
 paper (Stanghellini et al., in preparation).
 
 We have placed the central stars of the nebulae studied on the HR diagram and compared the central star locus with the
 evolutionary tracks for these post-AGB stars. We find strong systematics relating the dust type and
 the loci of their central stars on the HR diagram. For example, most F PNe are hosted by evolved central stars, 
 their
 dust features having probably disappeared due to the UV radiation field of their central stars. We also find that 
 central stars of CRD PNe are spread through the diagram, with the uncertain classified at low luminosity, and the 
 only SiC PNe for which we have the HR locus is in the early evolutionary stage for the central star as well.
  The last row of Table 4 clearly hints that
 CRD PNe have central stars that are typical for the galaxy they belong to 
 (see Villaver et al. 2006), while
 the only ORD PNe with measured mass is very massive. We need several
 more data points to substantiate this finding.
 
 Our observations show that there are fundamental
  differences in the characteristics of the grains formed around
  PNe at different metallicities, which are a consequence of the 
  more inefficient production of dust, the smaller mass-loss rates
  and slower evolution of the central star experienced by PNe at
  low metallicities compared to their Galactic analogs. At the low metallicities of the Magellanic Clouds
fewer grains are formed, with respect to their Galactic counterparts. The dust features that we see in our targets are
of moderate strengths.

\section{Conclusions}

The IRS spectra of a selection of Large and Small Magellanic Cloud PNe were analyzed for their dust content to
determine the nature of their dust chemistry and to gain knowledge of their dust evolutionary phase.
The spectra studied show three major dust emission profiles, and we classify then into:
F, those dominated by nebular emission lines; CRD, whose spectra show
carbonaceous dust; and ORD, with oxygen-rich dust compounds seen in emission.

The dust properties, including dust type, IR luminosity, and dust temperatures, do correlate strongly with the
known gas properties of the PNe. We find an exclusive correlation between dust type and carbon abundance, showing that
{\it all CRD PNe have C/O$>$1, and all ORD PNe have gone through the HBB phase}. We also find an exclusive 
correlation between CRD PNe and symmetric morphology, while all ORD PNe show high asymmetries. 
We also find hints of more massive central stars in ORD PNe.

The dust temperature decreases with increasing physical radii, and with decreasing electron density.
CRD PNe define a scattered evolutionary sequence, the scatter being due to the range of initial masses and
metallicity within the sample.

The statistics of dust emission features that we see in the Magellanic Clouds, if compared with 
that of the Galaxy, indicate that low metallicity environments do not favor the production of dust. As a consequence, 
the 
mass loss must be lower if the mass ejection is due to pressure on the dust grains. In particular, 
the spectacular bipolar 
nebulae, if indeed more massive and evolved from massive progenitors, are rarer in the Magellanic Clouds than in 
the galaxy, and the findings of the dust properties of this work conform to this scenario.

The PN sample analyzed in this paper represents an important fraction of all the LMC and SMC PNe ever
observed with {\it HST}. Further {\it HST} and Spitzer observations on a larger Magellanic 
Cloud PN sample, such as shown in the outer contours of Figure 1,  would be a good complement 
to this study. While the present Spitzer capabilities limit spectroscopic studies 
to bright Magellanic Cloud PNe, future space technology might 
extend the analysis to VLE PNe, such as those found in recent surveys 
(Reid \& Parker 2006ab). 

A comparative study of the LMC and SMC PNe together with those in the Galaxy, based on their mid-IR spectra, 
is also planned for the future.

\acknowledgments

Support for this work was provided by NASA through a grant issued by JPL/Caltech.
Arturo Manchado, Letizia Stanghellini, and Eva Villaver are grateful to the European Space 
Astronomy Centre for their hospitality during two science group meetings. 
Arturo Manchado acknowledges support from grant \emph{AYA 2004$-$3136}
from the Spanish Ministerio de Educaci\'on y Ciencia. We warmly thank Terry Mahoney for 
carefully reading the manuscript. We acknowledge the input from an anonymous referee.

\begin{deluxetable}{lrll}

\tablecaption{Target Selection}

\tablehead{
\colhead{Object} & 
\colhead{R [$\arcsec]$}& 
\colhead{Morphology}&
\colhead{notes} \\

}

\startdata
  
SMP~LMC~4 & 0.69& Elliptical& Faint halo\\
SMP~LMC~9 & 0.46& Bipolar core& Barrel shape\\
SMP~LMC~10 & 0.88& Pointsymmetric& Spiral shape\\
SMP~LMC~16 & 1.00& Bipolar& \\
SMP~LMC~18 & 0.51& Round& Possibly Bipolare core\\
SMP~LMC~19 & 0.41& Bipolar core& Ring\\
SMP~LMC~20 & 0.41& Bipolar core& \\ 
SMP~LMC~21 & 0.58& Quadrupolar&\\
SMP~LMC~25 & 0.23& Round&\\
SMP~LMC~27 & 0.46& Quadrupolar& Outer arc and halo\\
SMP~LMC~34 & 0.32& Elliptical& \\
SMP~LMC~45 & 0.72& Bipolar& Questionable morphology\\
SMP~LMC~46 & 0.31& Bipolar core& Possible ring\\
SMP~LMC~48 & 0.20& Elliptical& \\
SMP~LMC~66 & 0.54& Elliptical& \\
SMP~LMC~71 & 0.31& Elliptical&\\
SMP~LMC~72 & $\dots$    & Bipolar& \\
SMP~LMC~79 & 0.22& Bipolar core& \\
SMP~LMC~80 & 0.21& Round& \\
SMP~LMC~81 & 0.15& Round& Barely resolved\\
SMP~LMC~95 & 0.46& Bipolar core& Ansae\\
SMP~LMC~97 & 0.59& Round& \\ 
SMP~LMC~99 & 0.41& Bipolar core&\\ 
SMP~LMC~100 & 0.71& Bipolar core& Possibly quadrupolar\\
SMP~LMC~102 & 0.71& Round& Possibly Bipolar core\\

&&&\\
SMP~SMC~2 & 0.25& Round \\
SMP~SMC~5 & 0.31& Round \\ 
SMP~SMC~8 & 0.23&  Round&  \\
SMP~SMC~9 &  0.55& Round&  Inner structure \\
SMP~SMC~13 & 0.19&  Round& \\
SMP~SMC~14 & 0.42&  Round&  Ansae and/or inner structure\\
SMP~SMC~15 & 0.17& Round&\\
SMP~SMC~16 & 0.18& Elliptical& \\
SMP~SMC~17 & 0.25&  Elliptical& Faint detached halo \\
SMP~SMC~18 & 0.15&  $\dots$& Unresolved \\
SMP~SMC~19 & 0.30&  Round& Outer structure \\
SMP~SMC~20 & 0.15& $\dots$ & Unresolved \\
SMP~SMC~23 & 0.30&  Bipolar core&  \\
SMP~SMC~25 & 0.19& Elliptical&  \\
SMP~SMC~26 & 0.28& Pointsymmetric&  \\
SMP~SMC~27 & 0.23& Round& Attached outer halo\\
\\

\enddata 
\end{deluxetable}

\begin{deluxetable}{llllll}

\tablecaption{Observing log}

\tablehead{
\colhead{Object} & 
\colhead{R.A.}& 
\colhead{DEC}& 
\colhead{Obs. date} & 
\colhead{ID number} & 
\colhead{IRS campaign} \\

\colhead{}& 
\colhead{J2000.0}& 
\colhead{J2000.0} & 
\colhead{}& 
\colhead{}&
\colhead{}\\
}

\startdata
  
SMP~LMC~4 & 04$^{h}$43$^{m}$21$^{s}$.50 & $-$71$^{\rm o}$30$'$09$''$.5 & 08/14/2005 & 14703360 & 23.2	\\
SMP~LMC~9 & 04$^{h}$50$^{m}$24$^{s}$.71 & $-$68$^{\rm o}$13$'$17$''$.0 & 09/13/2005 & 14705920 & 24	\\
SMP~LMC~10 & 04$^{h}$51$^{m}$08$^{s}$.90 & $-$68$^{\rm o}$49$'$05$''$.8 & 09/13/2005 & 14700032 & 24	\\
SMP~LMC~16 & 05$^{h}$02$^{m}$01$^{s}$.91 & $-$69$^{\rm o}$48$'$54$''$.4 & 08/14/2005 & 14700288 & 23.2	\\
SMP~LMC~18 & 05$^{h}$03$^{m}$42$^{s}$.64 & $-$70$^{\rm o}$06$'$47$''$.8 & 08/14/2005 & 14700544 & 23.2	\\
SMP~LMC~19 & 05$^{h}$03$^{m}$41$^{s}$.30 & $-$70$^{\rm o}$13$'$53$''$.6 & 08/14/2005 & 14700800 & 23.2	\\
SMP~LMC~20 & 05$^{h}$04$^{m}$40$^{s}$.14 & $-$69$^{\rm o}$21$'$39$''$.3 & 08/14/2005 & 14701056 & 23.2	\\
SMP~LMC~21 & 05$^{h}$04$^{m}$51$^{s}$.90 & $-$68$^{\rm o}$39$'$10$''$.0 & 08/14/2005 & 14701312 & 23.2	\\
SMP~LMC~25 & 05$^{h}$06$^{m}$24$^{s}$.00 & $-$69$^{\rm o}$03$'$19$''$.2 & 08/14/2005 & 14701568 & 23.2	\\
SMP~LMC~27 & 05$^{h}$07$^{m}$54$^{s}$.90 & $-$66$^{\rm o}$57$'$46$''$.1 & 09/13/2005 & 14701824 & 24	\\
SMP~LMC~34 & 05$^{h}$10$^{m}$17$^{s}$.18 & $-$68$^{\rm o}$48$'$23$''$.0 & 08/14/2005 & 14702336 & 23.2	\\
SMP~LMC~45 & 05$^{h}$19$^{m}$21$^{s}$.00 & $-$66$^{\rm o}$58$'$13$''$.0 & 07/14/2005 & 14702592 & 22	\\
SMP~LMC~46 & 05$^{h}$19$^{m}$29$^{s}$.72 & $-$68$^{\rm o}$51$'$09$''$.1 & 08/11/2005 & 14702848 & 23.2	\\
SMP~LMC~48 & 05$^{h}$20$^{m}$09$^{s}$.66 & $-$69$^{\rm o}$53$'$39$''$.2 & 08/14/2005 & 14703104 & 23.2	\\
SMP~LMC~66 & 05$^{h}$28$^{m}$41$^{s}$.20 & $-$67$^{\rm o}$33$'$39$''$.0 & 08/11/2005 & 14703616 & 23.2	\\
SMP~LMC~71 & 05$^{h}$30$^{m}$33$^{s}$.22 & $-$70$^{\rm o}$44$'$38$''$.4 & 08/11/2005 & 14703872 & 23.2	\\
SMP~LMC~72 & 05$^{h}$30$^{m}$45$^{s}$.98 & $-$70$^{\rm o}$50$'$16$''$.4 & 08/11/2005 & 14704128 & 23.2	\\
SMP~LMC~79 & 05$^{h}$34$^{m}$08$^{s}$.76 & $-$74$^{\rm o}$20$'$06$''$.6 & 10/10/2005 & 14704384 & 25	\\
SMP~LMC~80 & 05$^{h}$34$^{m}$38$^{s}$.87 & $-$70$^{\rm o}$19$'$56$''$.9 & 08/11/2005 & 14704640 & 23.2	\\
SMP~LMC~81 & 05$^{h}$35$^{m}$20$^{s}$.92 & $-$73$^{\rm o}$55$'$30$''$.1 & 10/10/2005 & 14704896 & 25	\\
SMP~LMC~95 & 06$^{h}$01$^{m}$45$^{s}$.30 & $-$67$^{\rm o}$56$'$08$''$.0 & 08/11/2005 & 14705152 & 23.2	\\
SMP~LMC~97 & 06$^{h}$10$^{m}$25$^{s}$.50 & $-$67$^{\rm o}$56$'$21$''$.0 & 08/11/2005 & 14705408 & 23.2	\\
SMP~LMC~99 & 06$^{h}$18$^{m}$58$^{s}$.21 & $-$71$^{\rm o}$35$'$50$''$.7 & 10/15/2005 & 14705664 & 25	\\
SMP~LMC~100 & 06$^{h}$22$^{m}$55$^{s}$.73 & $-$72$^{\rm o}$07$'$41$''$.4 & 11/14/2005 & 14699520 & 26	\\
SMP~LMC~102 & 06$^{h}$29$^{m}$32$^{s}$.93 & $-$68$^{\rm o}$03$'$32$''$.9 & 08/11/2005 & 14699776 & 23.2	\\

&&&&&\\
SMP~SMC~2 & 00$^{h}$32$^{m}$38$^{s}$.81 & $-$71$^{\rm o}$41$'$58$''$.7 & 08/14/2005 & 14709248 & 23.2	\\
SMP~SMC~5 & 00$^{h}$41$^{m}$21$^{s}$.67 & $-$72$^{\rm o}$45$'$18$''$.0 & 08/14/2005 & 14709504 & 23.2	\\
SMP~SMC~8 & 00$^{h}$43$^{m}$25$^{s}$.17 & $-$72$^{\rm o}$38$'$18$''$.9 & 08/14/2005 & 14709760 & 23.2	\\
SMP~SMC~9 & 00$^{h}$45$^{m}$20$^{s}$.66 & $-$73$^{\rm o}$24$'$10$''$.5 & 08/14/2005 & 14710016 & 23.2	\\
SMP~SMC~13 & 00$^{h}$49$^{m}$51$^{s}$.71 & $-$73$^{\rm o}$44$'$21$''$.3 & 08/14/2005 & 14706176 & 23.2	\\
SMP~SMC~14 & 00$^{h}$50$^{m}$34$^{s}$.99 & $-$73$^{\rm o}$42$'$57$''$.9 & 08/14/2005 & 14706432 & 23.2	\\
SMP~SMC~15 & 00$^{h}$51$^{m}$07$^{s}$.45 & $-$73$^{\rm o}$57$'$37$''$.1 & 08/14/2005 & 14706688 & 23.2	\\
SMP~SMC~16 & 00$^{h}$51$^{m}$27$^{s}$.08 & $-$72$^{\rm o}$26$'$11$''$.1 & 08/14/2005 & 14706944 & 23.2	\\
SMP~SMC~17 & 00$^{h}$51$^{m}$56$^{s}$.44 & $-$71$^{\rm o}$24$'$44$''$.1 & 08/08/2005 & 14707200 & 23.2	\\
SMP~SMC~18 & 00$^{h}$51$^{m}$57$^{s}$.97 & $-$73$^{\rm o}$20$'$30$''$.1 & 08/14/2005 & 14707456 & 23.2	\\
SMP~SMC~19 & 00$^{h}$53$^{m}$11$^{s}$.14 & $-$72$^{\rm o}$45$'$07$''$.5 & 08/14/2005 & 14707712 & 23.2	\\
SMP~SMC~20 & 00$^{h}$56$^{m}$05$^{s}$.39 & $-$70$^{\rm o}$19$'$24$''$.7 & 08/08/2005 & 14707968 & 23.2	\\
SMP~SMC~23 & 00$^{h}$58$^{m}$42$^{s}$.14 & $-$72$^{\rm o}$56$'$59$''$.6 & 08/14/2005 & 14708224 & 23.2	\\
SMP~SMC~25 & 00$^{h}$59$^{m}$40$^{s}$.51 & $-$71$^{\rm o}$38$'$15$''$.3 & 08/14/2005 & 14708480 & 23.2	\\
SMP~SMC~26 & 01$^{h}$04$^{m}$17$^{s}$.81 & $-$73$^{\rm o}$21$'$51$''$.2 & 08/14/2005 & 14708736 & 23.2	\\
SMP~SMC~27 & 01$^{h}$21$^{m}$10$^{s}$.67 & $-$73$^{\rm o}$14$'$35$''$.4 & 08/14/2005 & 14708992 & 23.2	\\

\enddata 
\end{deluxetable}

\begin{deluxetable}{lrcrll}

\tablecaption{Characteristics of the IRS spectra}

\tablehead{
\colhead{Object} & 
\colhead{T$_{\rm dust}$}& 
\colhead{log F$_{\rm IR}$}& 
\colhead{IRE} & 
\colhead{Dust type} & 
\colhead{Excitation} \\

\colhead{}& 
\colhead{[K]}& 
\colhead{[erg cm$^{-2}$ s$^{-1}$]} & 
\colhead{}& 
\colhead{}&
\colhead{}\\
}

\startdata
  
      SMP~LMC~4 & 190 & -11.7 & 2.65 & F & very high \\ 
       SMP~LMC~9 & 140 & -11.7 & 0.86 & CRD & very high \\ 
       SMP~LMC~10 & 90 & -12.5 & 0.10 & F & high \\ 
       SMP~LMC~16 & 80 & -12.3 & 0.39 & F & very high \\ 
       SMP~LMC~18 & 110 & -11.9 & 1.2 & F & intermediate \\ 
       SMP~LMC~19 & 130 & -11.3 & 1.05 & CRD & very high \\ 
       SMP~LMC~20 & 90& -12.3 & 0.24 & F & very high \\ 
       SMP~LMC~21 & 130 & -11.1 & 20.96 & ORD & very high \\ 
       SMP~LMC~25 & 160 & -10.7 & 1.52 & CRD & intermediate \\ 
       SMP~LMC~27 & 140 & -11.9 & 1.75 & F & intermediate \\ 
       SMP~LMC~34 & 130 & -11.4 & 1.40 & F & intermediate \\ 
       SMP~LMC~45 & 120 & -11.9 & 0.30 & F & high \\ 
       SMP~LMC~46 & 140 & -11.9 & 1.37 & CRD & high \\ 
       SMP~LMC~48 & 170 & -10.8 & 1.43 & CRD & intermediate \\ 
       SMP~LMC~66 & 140 & -11.6 & 0.96 & F & high \\ 
       SMP~LMC~71 & 150 & -10.9 & 2.72 & CRD & very high \\ 
       SMP~LMC~72 & 60 & -13.0 & 1.88 & F & very high \\ 
       SMP~LMC~79 & 170 & -10.9 & 2.64 & CRD & high \\ 
       SMP~LMC~80 & 110 & -13.1 & 0.05 & F & intermediate \\ 
       SMP~LMC~81 & 130 & -10.6 & 2.79 & ORD & intermediate \\ 
       SMP~LMC~95 & 110 & -12.1 & 0.68 & F & high \\ 
       SMP~LMC~97 & 130 & -11.7 & 0.20 & F & very high \\ 
       SMP~LMC~99 & 160 & -11.0 & 0.68 & CRD & very high \\ 
      SMP~LMC~100 & 120 & -11.5 & 0.74 & CRD & very high \\ 
      SMP~LMC~102 & 140 & -12.1 & 0.62 & F & high \\

&&&&&\\

       SMP~SMC~2 & 160 & -11.5 & 0.79 & CRD & very high \\ 
       SMP~SMC~5 & 180 & -11.1 & $\dots$ & CRD & very high \\ 
       SMP~SMC~8 & 160 & -11.7 & 0.49 & F & intermediate \\ 
       SMP~SMC~9 & 120 & -12.0 & 1.10 & F & $\dots$ \\ 
       SMP~SMC~13 & 190 & -11.3 & 0.60 & CRD & intermediate \\ 
       SMP~SMC~14 & 150 & -11.6 & 0.91 & CRD & high \\ 
       SMP~SMC~15 & 190 & -10.8 & 2.03 & CRD & intermediate \\ 
       SMP~SMC~16 & 180 & -11.4 & 0.87 & CRD & low \\ 
       SMP~SMC~17 & 160 & -11.1 & 1.19 & CRD & intermediate \\ 
       SMP~SMC~18 & 170 & -10.8 & 2.37 & CRD & intermediate \\ 
       SMP~SMC~19 & 150 & -11.3 & 1.50 & CRD & very high \\ 
       SMP~SMC~20 & 250 & -11.0 & 1.22 & CRD & intermediate \\ 
       SMP~SMC~23 & 150 & -12.2 & 0.35 & F & intermediate \\ 
       SMP~SMC~25 & 130 & -11.5 & 1.86 & ORD & very high \\ 
       SMP~SMC~26 & 130 & -12.1 & 0.76 & F & very high \\ 
       SMP~SMC~27 & 180 & -11.4 & 0.47 & CRD & intermediate \\

\enddata 
\end{deluxetable} 

\begin{deluxetable}{lrrrrrrrr}
\rotate
\tablewidth{20truecm}
\tablecaption{Summary of relations}

\tablehead{

\colhead{}&
\colhead{}&
\multicolumn{3}{c}{LMC}&
\colhead{}&
\multicolumn{3}{c}{SMC}\\
\cline{3-5}
\cline{7-9}
\\

\colhead{}&
\colhead{}&
\colhead{F}&
\colhead{CRD}&
\colhead{ORD}&
\colhead{}&
\colhead{F}&
\colhead{CRD}&
\colhead{ORD}\\

}

\startdata
  
Sample size&	 &	14&	9&	2&	& 4&	11&	1\\

$<$C/H$>\times10^4$&	&	$(2.0\pm2.0)$& $(4.3\pm2.0)$& $(0.18\pm0.05)$& & $(4.2\pm2.7)$& $(4.6\pm2.0)$& $\dots$ \\

$<$N/H$>\times10^5$&	& $(8.4\pm0.1)$& $(7.6\pm4.1)$& $(12\pm15)$& & $(7.4\pm7.4)$& $(2.6\pm1.7)$& 8.3\\

$<$O/H$>\times10^4$&	&	$(1.8\pm1.0)$& $(2.4\pm1.0)$& $(1.3\pm0.7)$& & $(2.1\pm1.7)$& $(1.4\pm0.6)$& 0.4\\

$<$Ne/H$>\times10^5$&	&	$(2.7\pm1.5)$& $(4.1\pm1.9)$& $(1.8\pm0.4)$& & $(3.4\pm4.9)$& $(2.2\pm1.6)$& 0.9\\

$<$M/M$_{\odot}>$&	& $(0.65\pm0.13)$& $(0.63\pm0.00)$& $\dots$&& 0.59& $(0.59\pm0.05)$& 0.82\\

\enddata 
\end{deluxetable} 

\clearpage

\begin{figure}
\figurenum{1}
\plotone{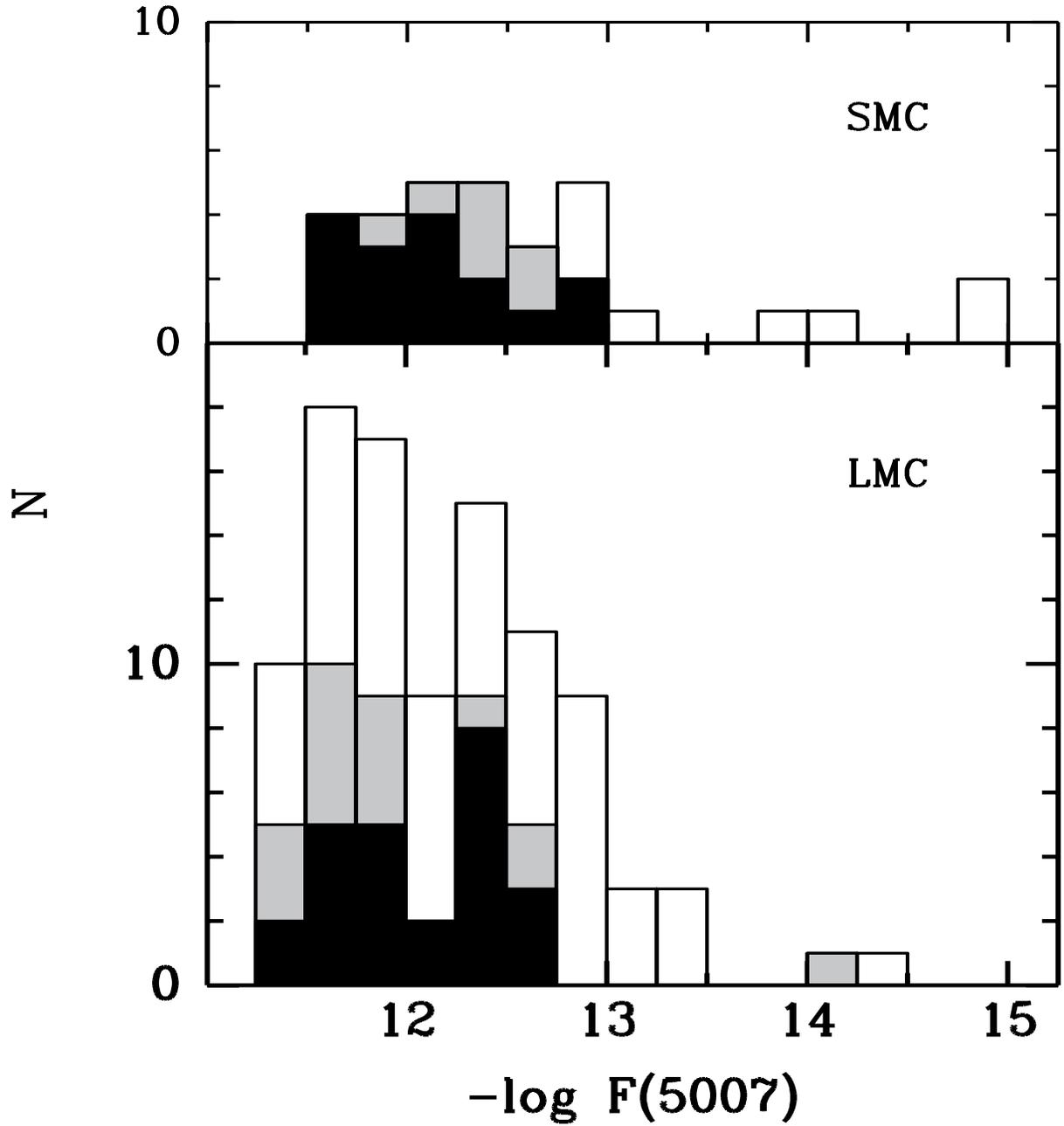}
\caption{Histogram of the [\ion{O}{3}] $\lambda$5007 flux of
Magellanic Cloud PNe whose size is smaller
than 2 $\arcsec$ as measured on the {\it HST} images. Outer histogram: complete sample. Gray histogram:
the ROC sample (i.e., the GTO program by Houck). Black histogram: the PN sample presented in this paper.}
\end{figure}

\begin{figure}
\figurenum{2}
\plotone{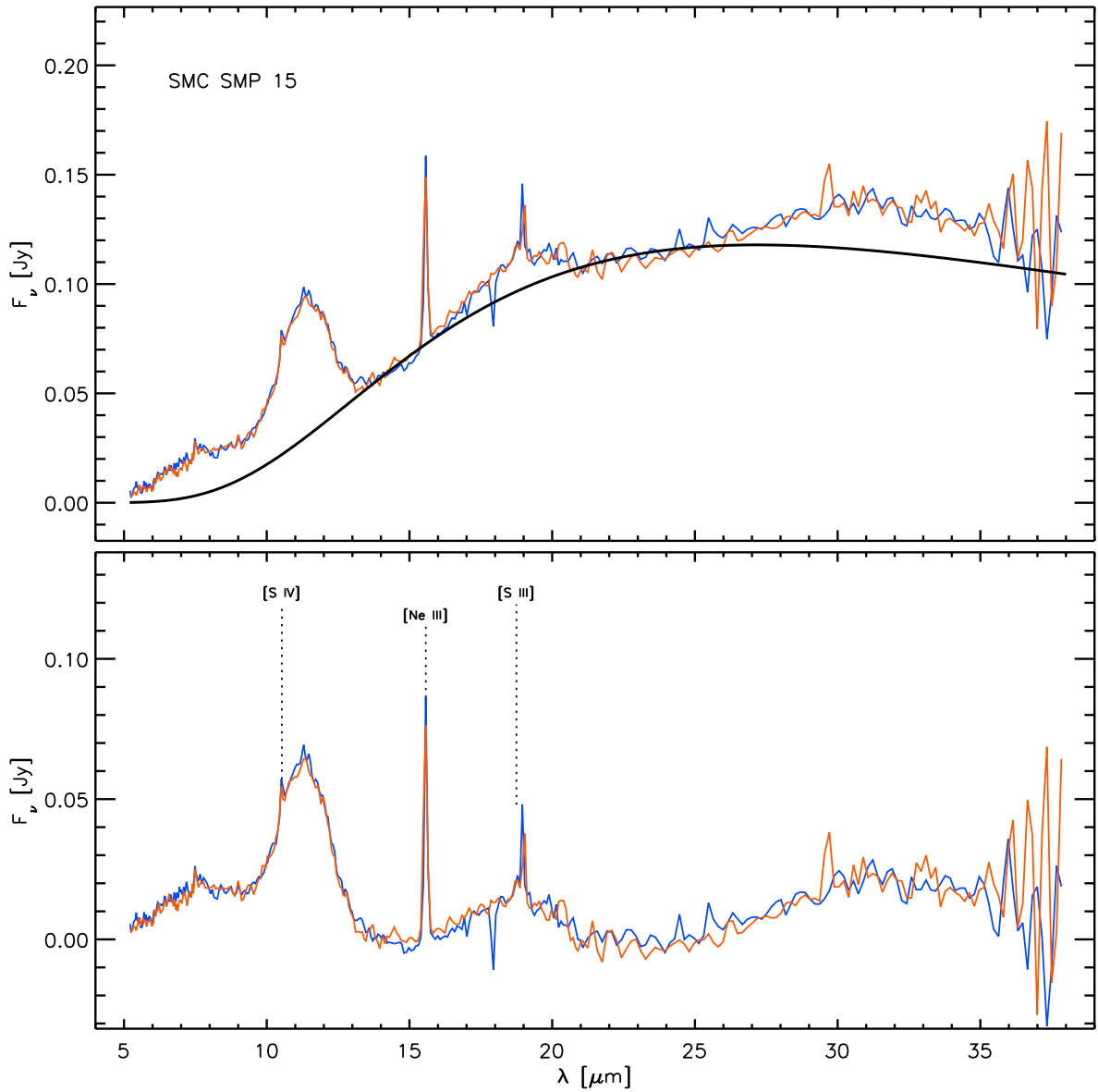}
\caption{Continuum fitted (upper panel) and continuum-subtracted spectrum (lower panel)
of SMP~SMC~15. Both nod positions are shown, in thin lines. The thick
line in the upper plot identifies the blackbody fit used. Note that the fit is correctly
below the broad dust features: the large silicon carbide feature at 11 $\mu$m, 
the broad feature at 15--21 $\mu$m, and the broad feature peaking at 30 $\mu$m.
Several nebular emission lines have
been identified in the lower panel. The IRS spectral type of this PN is CRD (see text of $\S$2.3
for spectral type definitions).
}
\end{figure}

\begin{figure}
\figurenum{3}
\plotone{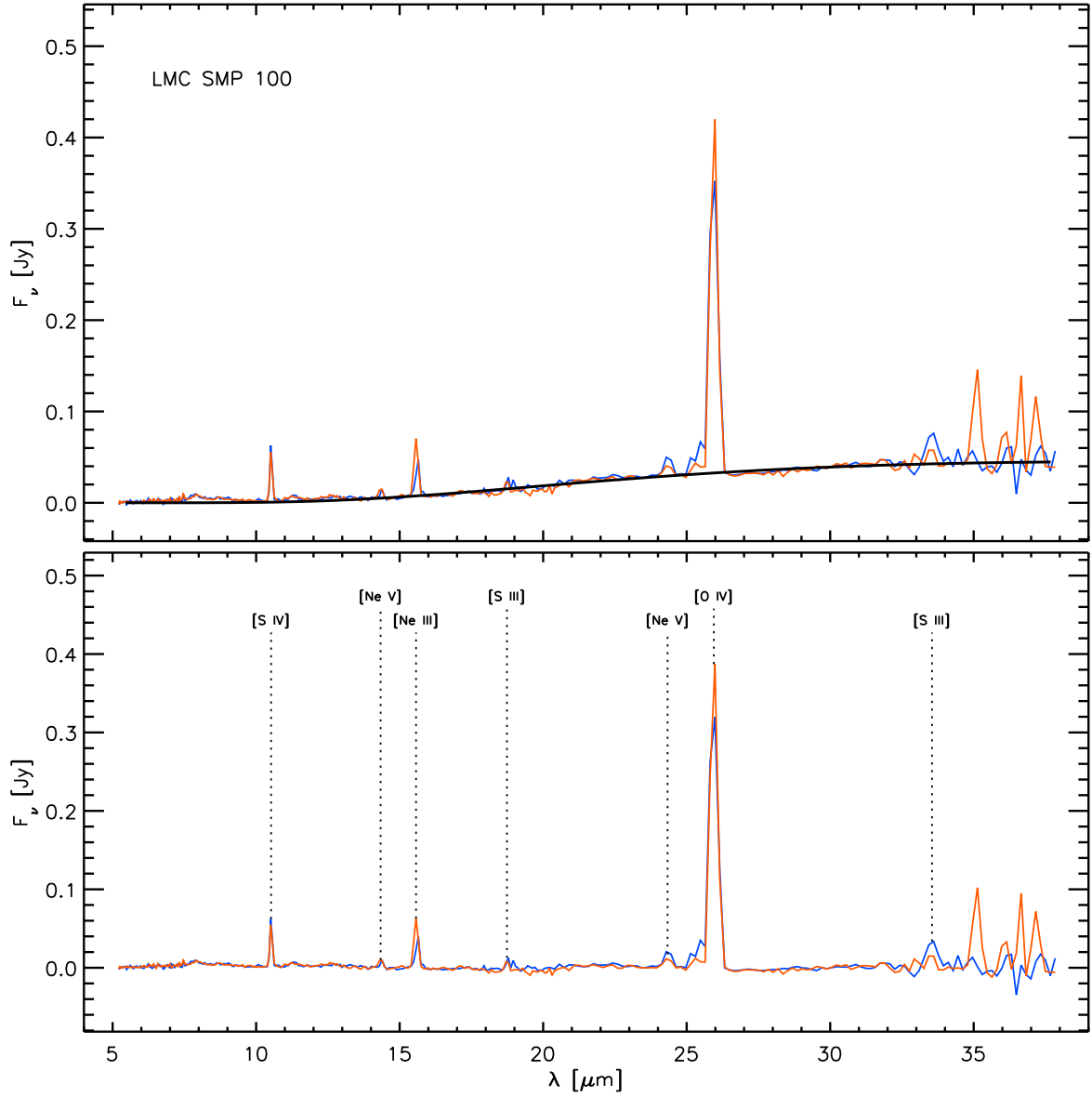}
\caption{Same as in Fig.~2, for SMP~LMC~100. The spectrum is dominated by the emission lines with 
only a faint carbonaceous dust features at 6.2 $\mu$m. The IRS spectral type of this PN is CRD.
}
\end{figure}

\begin{figure}
\figurenum{4}
\plotone{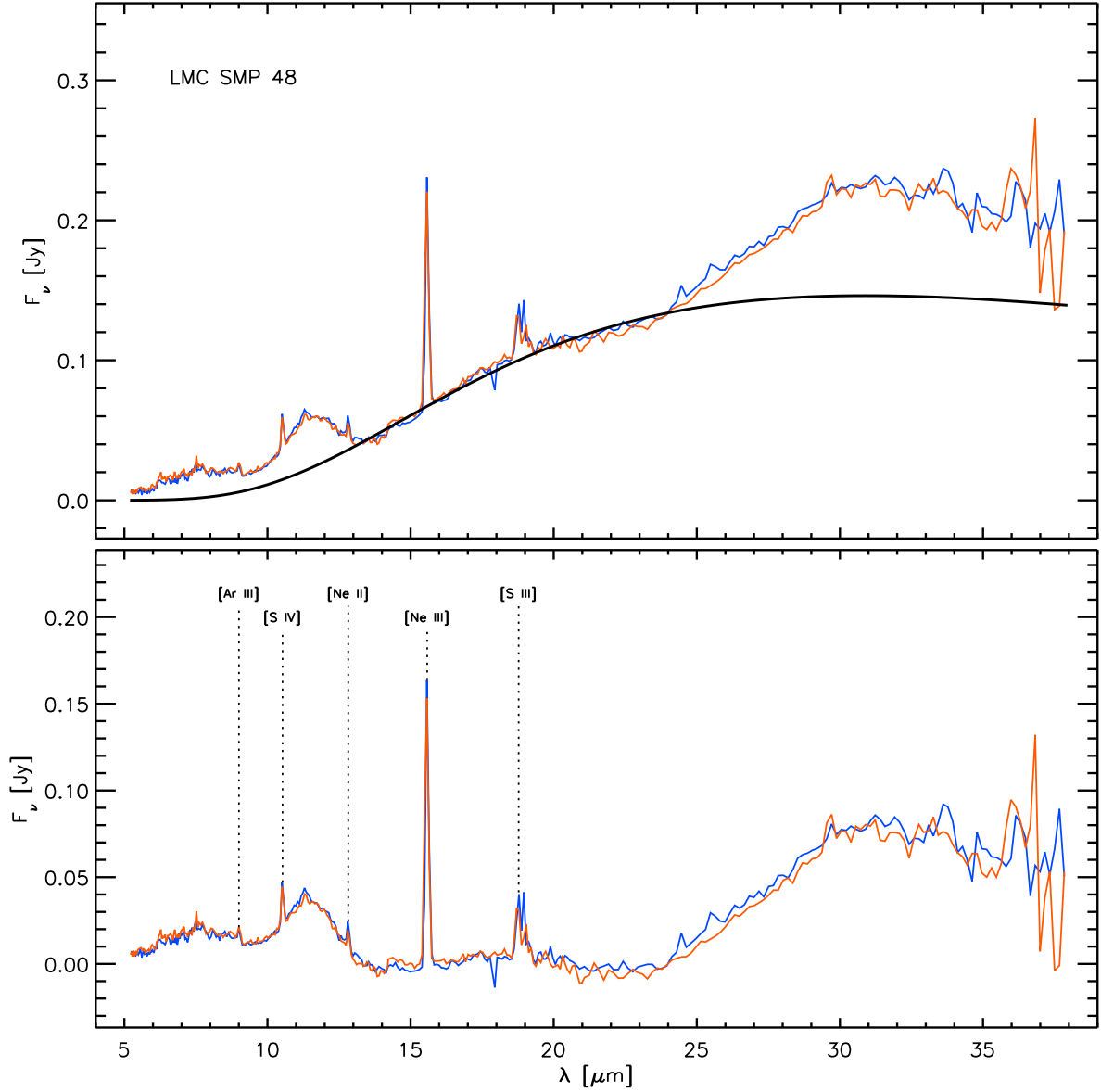}
\caption{Same as in Fig.~2, for SMP~LMC~48. The broad plateau feature at 6--9 $\mu$m and the renmant of a broad
amorphous SiC feature at 11 $\mu$m are clearly detected in this spectrum, together with a very strong 30 $\mu$m feature.
The IRS spectral type of this PN is CRD.}
\end{figure}

\begin{figure}
\figurenum{5}
\plotone{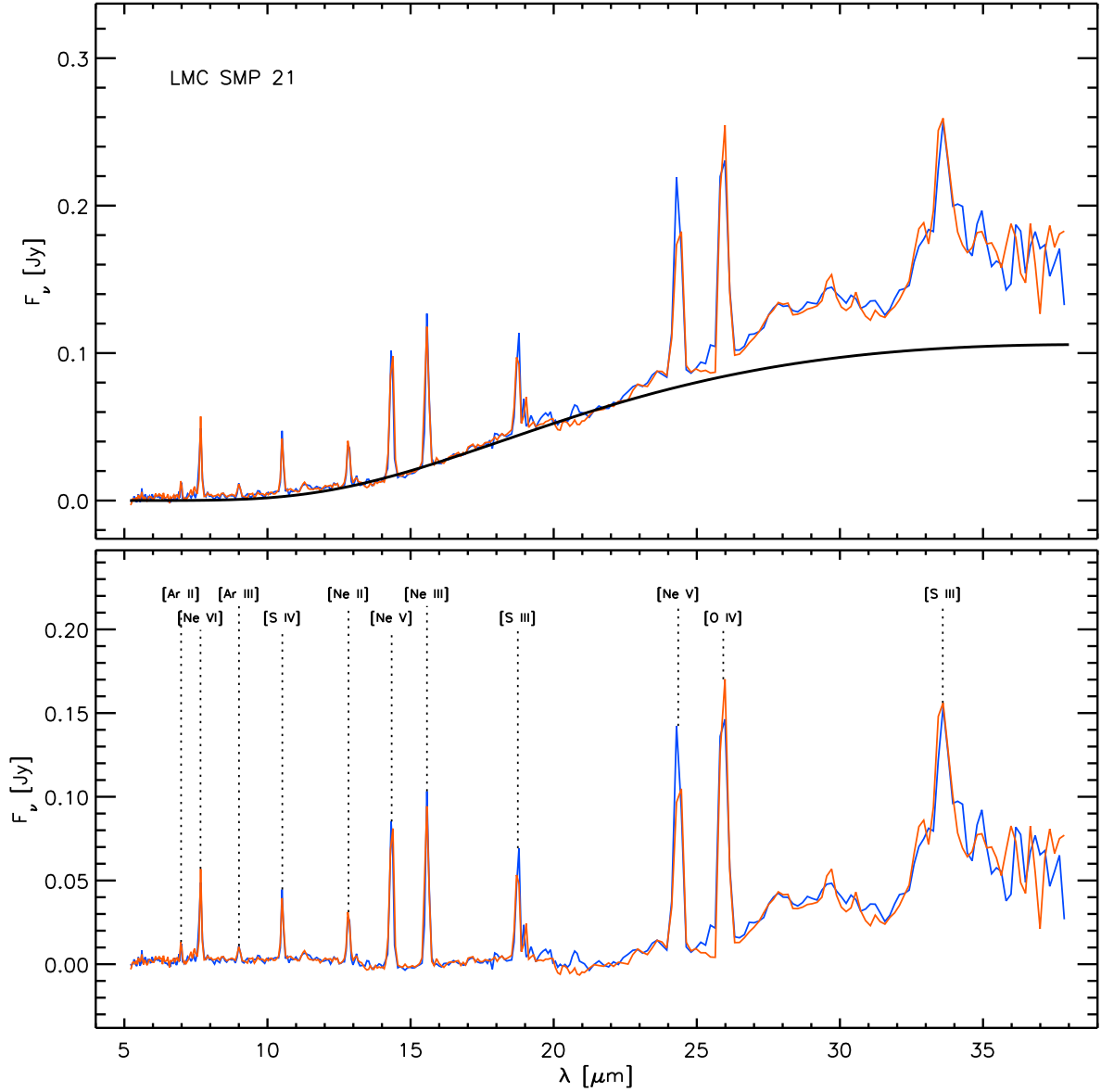}
\caption{Same as in Fig.~2, for SMP~LMC~21. The oxygen-rich dust features are seen 
at longer wavelengths. The IRS spectral type of this PN is ORD.}
\end{figure}

\clearpage
\begin{figure}
\epsscale{.8}
\figurenum{6a}
\plotone{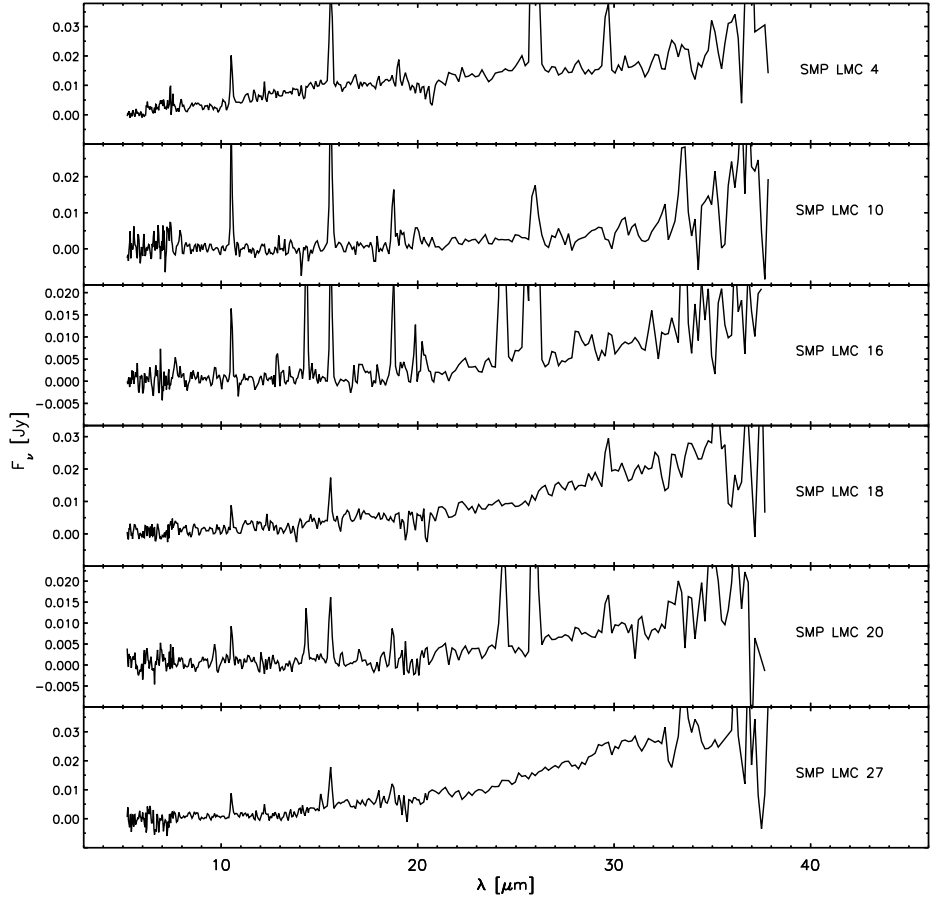}
\caption{IRS spectra of F PNe.}
\end{figure}
\clearpage
{\plotone{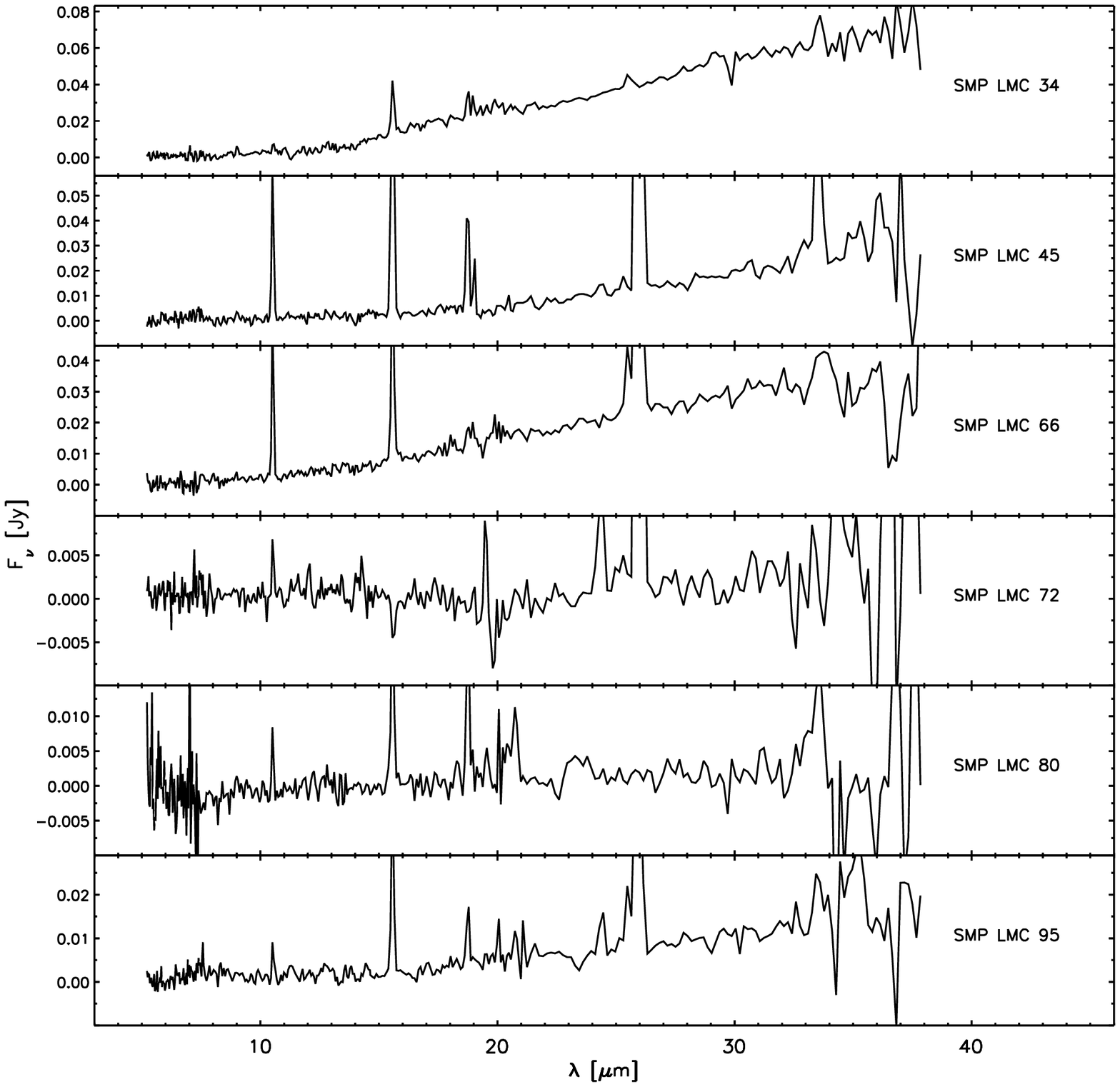}}\\
\centerline{Fig. 6b. --- Continued.}
\clearpage
{\plotone{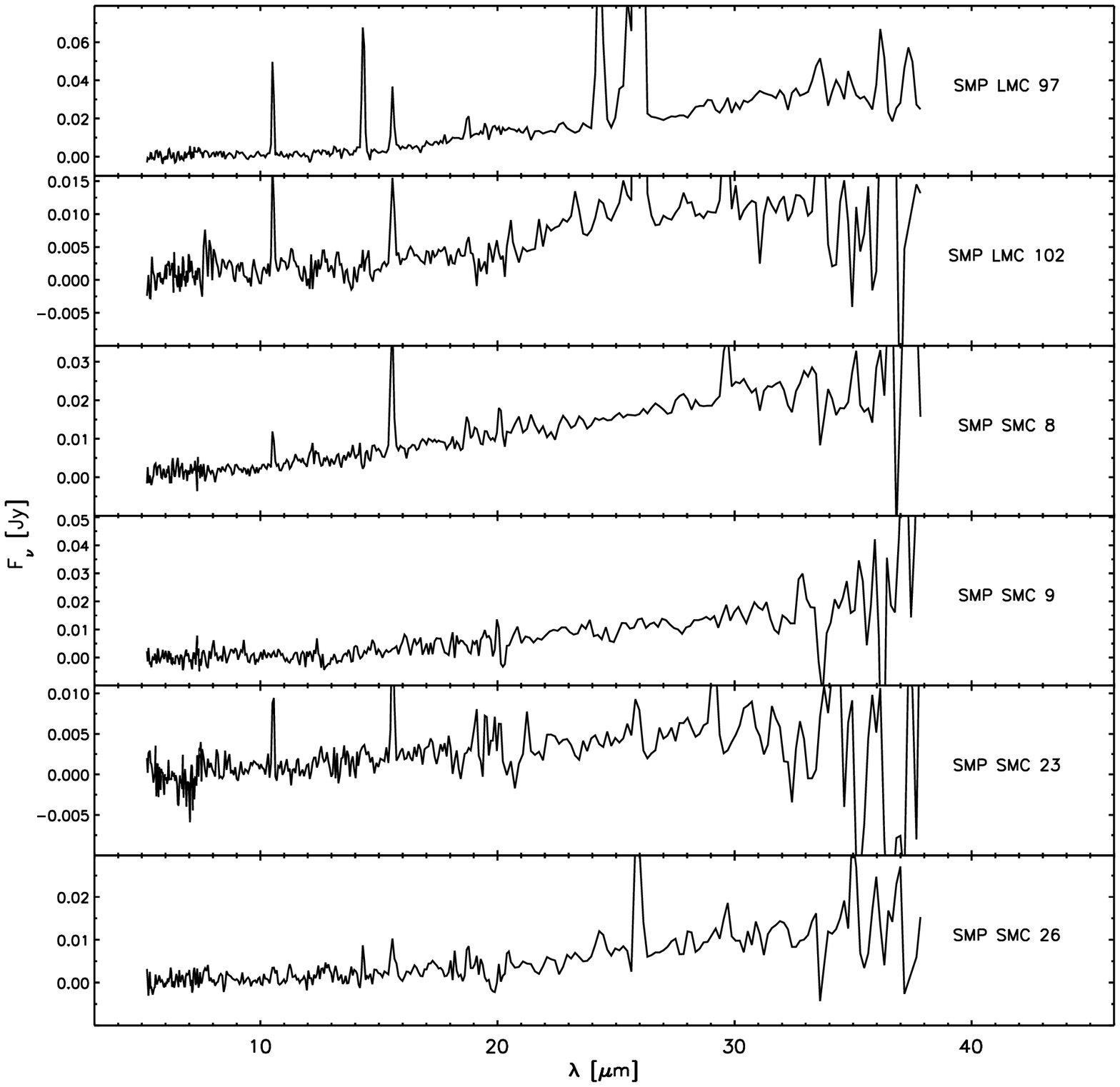}}\\
\centerline{Fig. 6c. --- Contiunued.}
\clearpage

\begin{figure}
\figurenum{7a}
\plotone{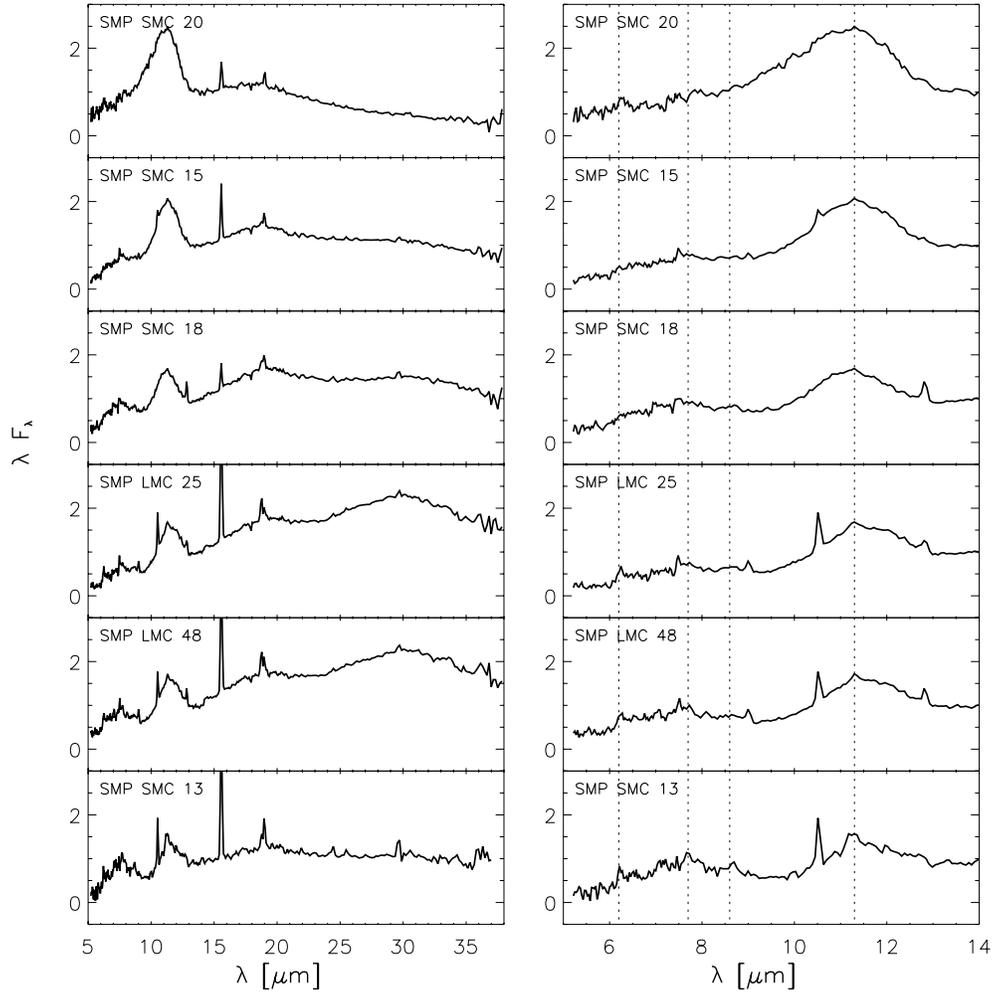}
\caption{IRS spectra of CRD PNe. Left panel: entire spectra. Right panels: the 5--14 $\mu$m spectra, where the vertical
lines correspond to the classical PAH features at 6.2, 7.7, 8.6, and 11.3 $\mu$m.}
\end{figure}
\clearpage
{\plotone{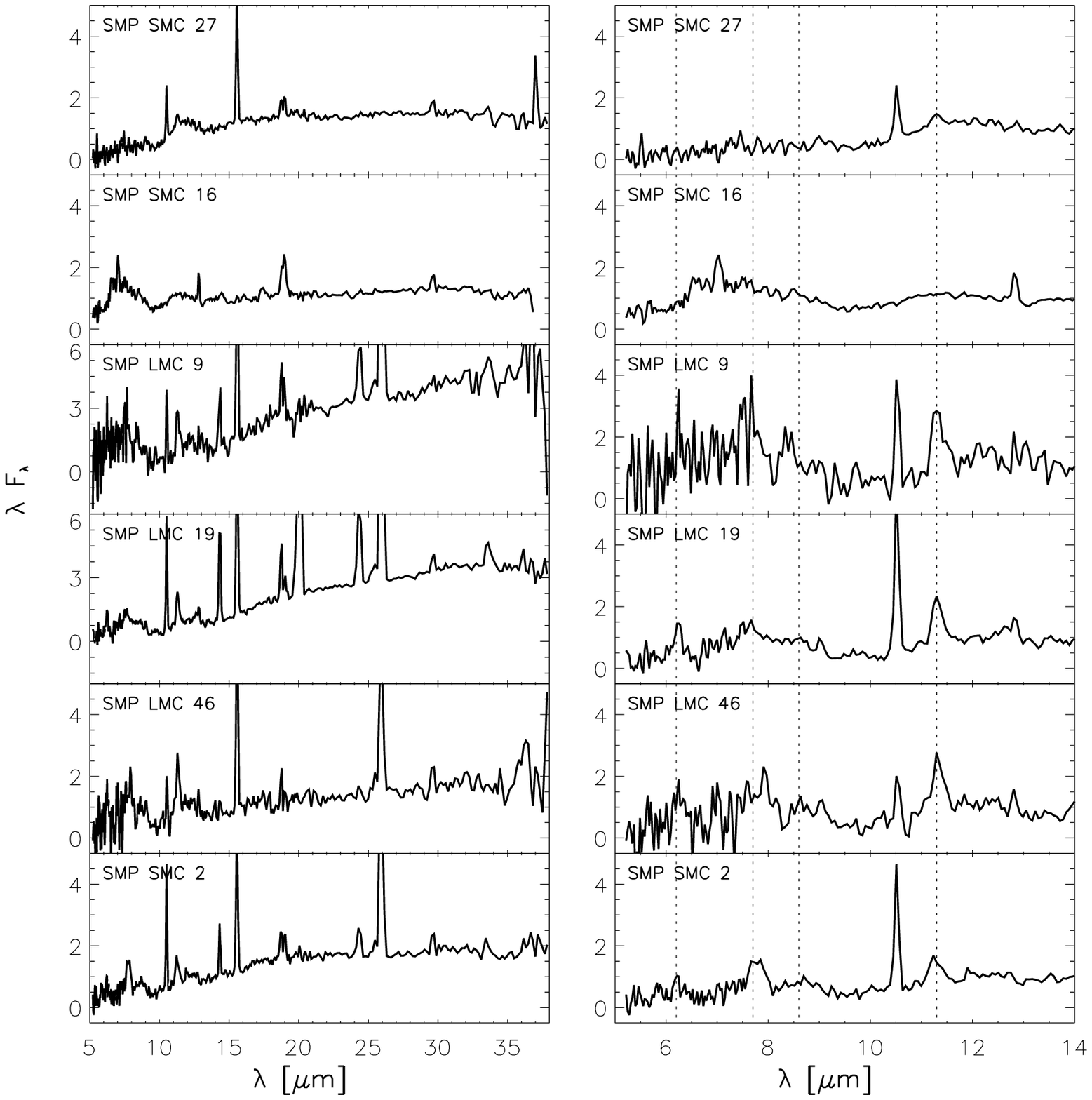}}\\
\centerline{Fig. 7b. --- Continued.}
\clearpage
{\plotone{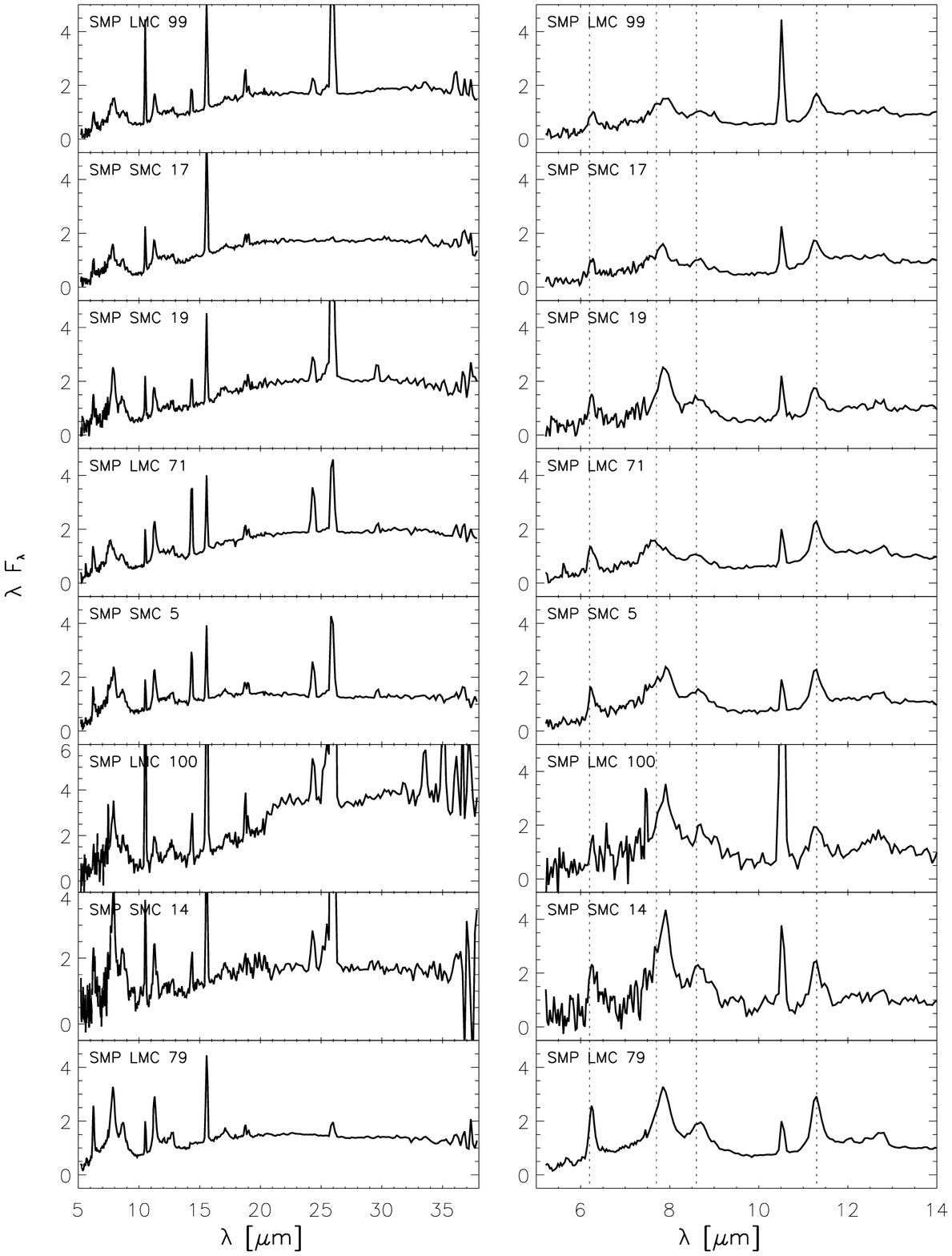}}\\
\centerline{Fig. 7c. --- Continued.}
\clearpage

\begin{figure}
\figurenum{8}
\epsscale{1}
\plotone{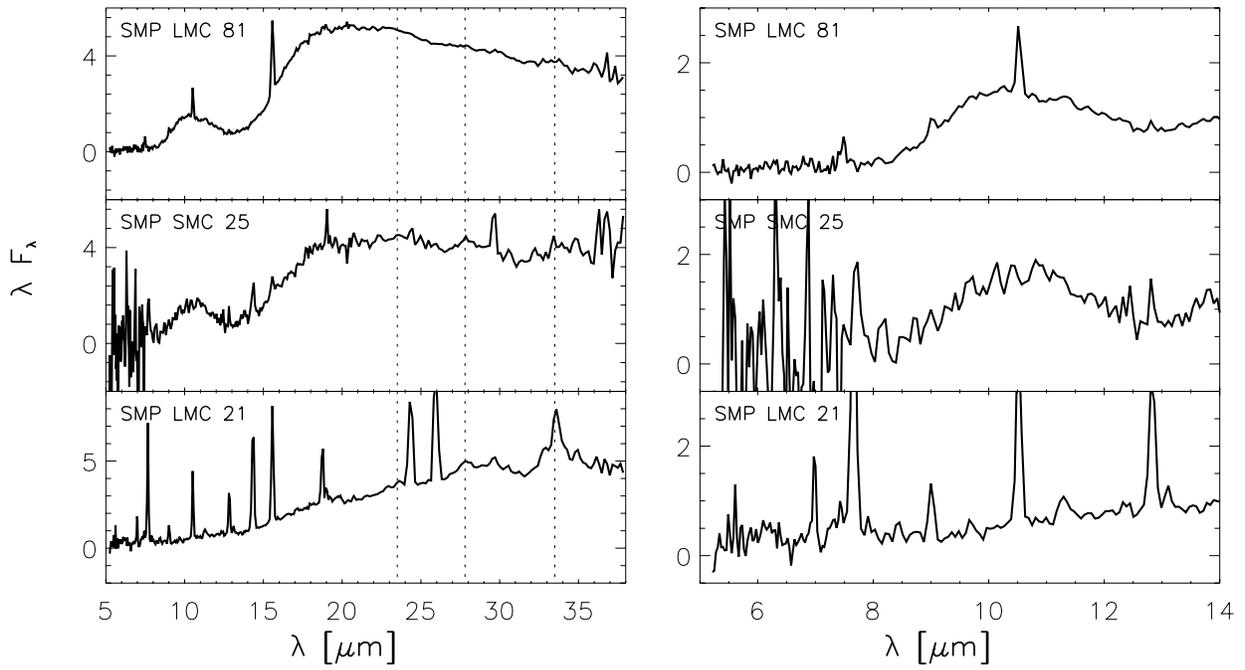}
\caption{IRS spectra of ORD PNe. Left panel: entire spectra, where the vertical lines correspond
to the crystalline silicate features at 23.5, 27.8, and 33.5 $\mu$m. Right panel: the 5--14 $\mu$m spectra.}
\end{figure}

 \begin{figure}
\figurenum{9}
\includegraphics[bb = 100 100 600 600, angle=0]{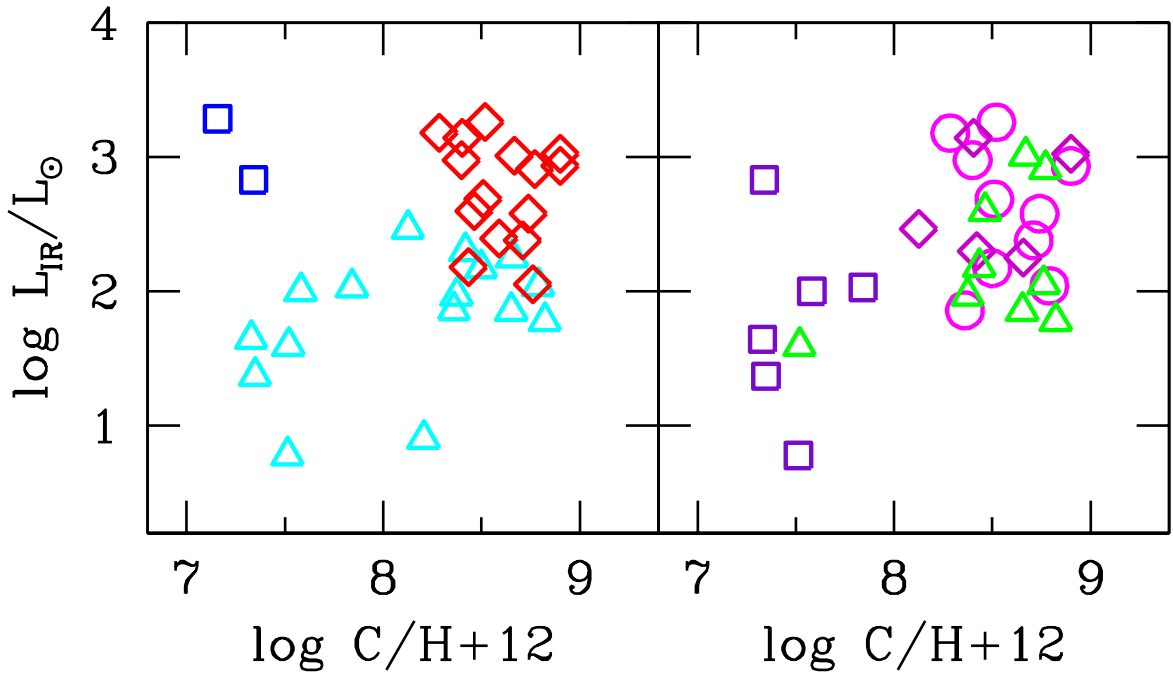}
\caption{Infrared luminosity versus carbon abundance. Left panel:
dust type is coded with 
triangles (F), diamonds (CRD), and squares (ORD). 
Right panel: morphology is coded with circles (Round), diamonds (Elliptical), triangles
(Bipolar Core), and squares (Bipolar, Quadrupolar, and Pointsymmetric).}
\end{figure}

\begin{figure}
\figurenum{10}
\includegraphics[bb = 100 100 600 600, angle=0]{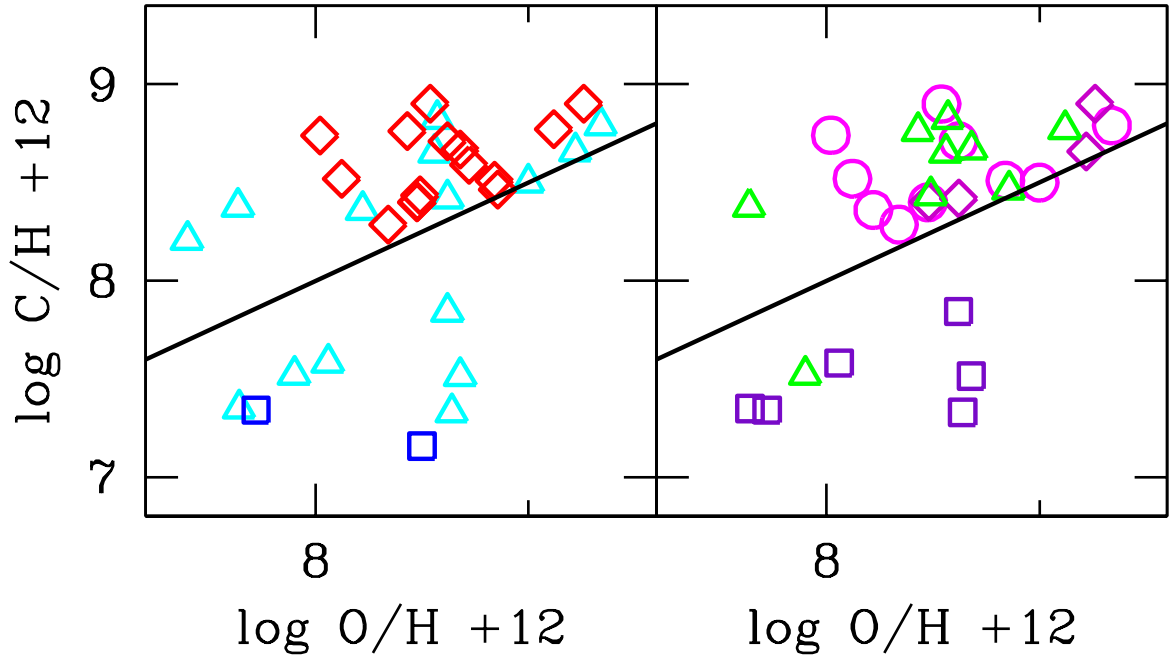}
\caption{Carbon versus oxygen abundance of the gas of Magellanic Cloud PNe. Dust type (left panel)
and morphology (right panel) are coded as in Fig.~9. The line represents C/O=1.}
\end{figure}

\clearpage

\begin{figure}
\figurenum{11}
\includegraphics[bb = 100 100 600 600, angle=0]{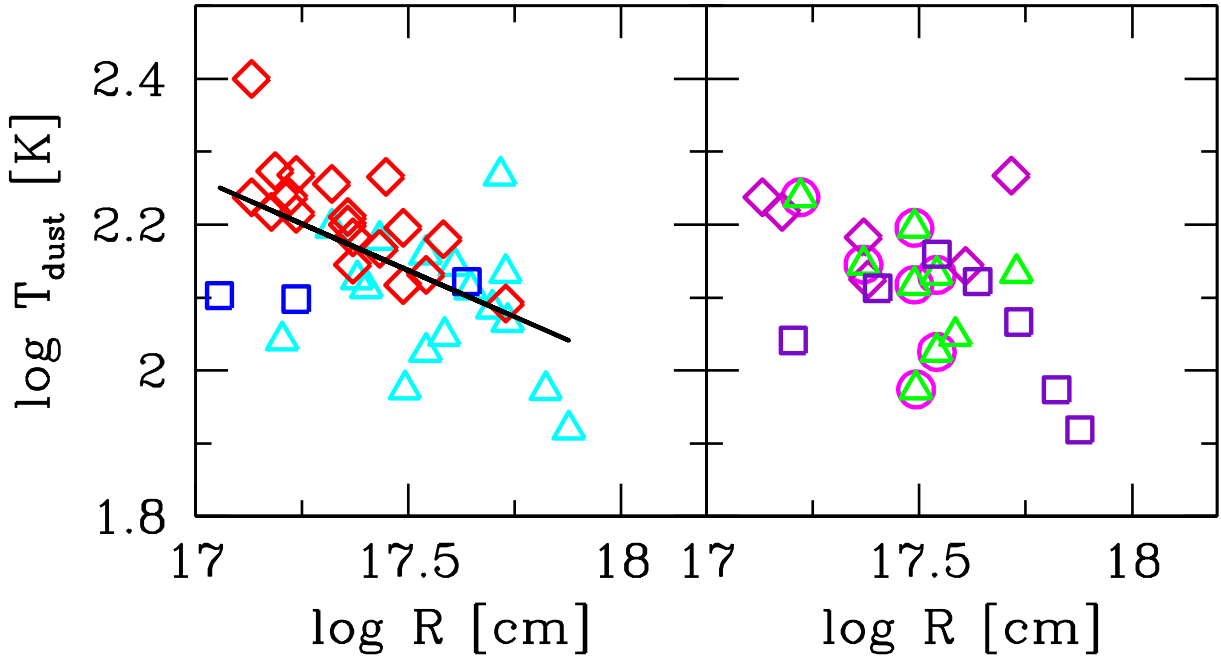}
\caption{The relation between T$_{\rm dust}$ and physical radius. Symbols for dust type (left panel)
and morphology (right panel) are coded as in Fig.~9. The line is the best fit to the data. }
\end{figure}

\begin{figure}
\figurenum{12}
\includegraphics[bb = 100 100 600 600, angle=0]{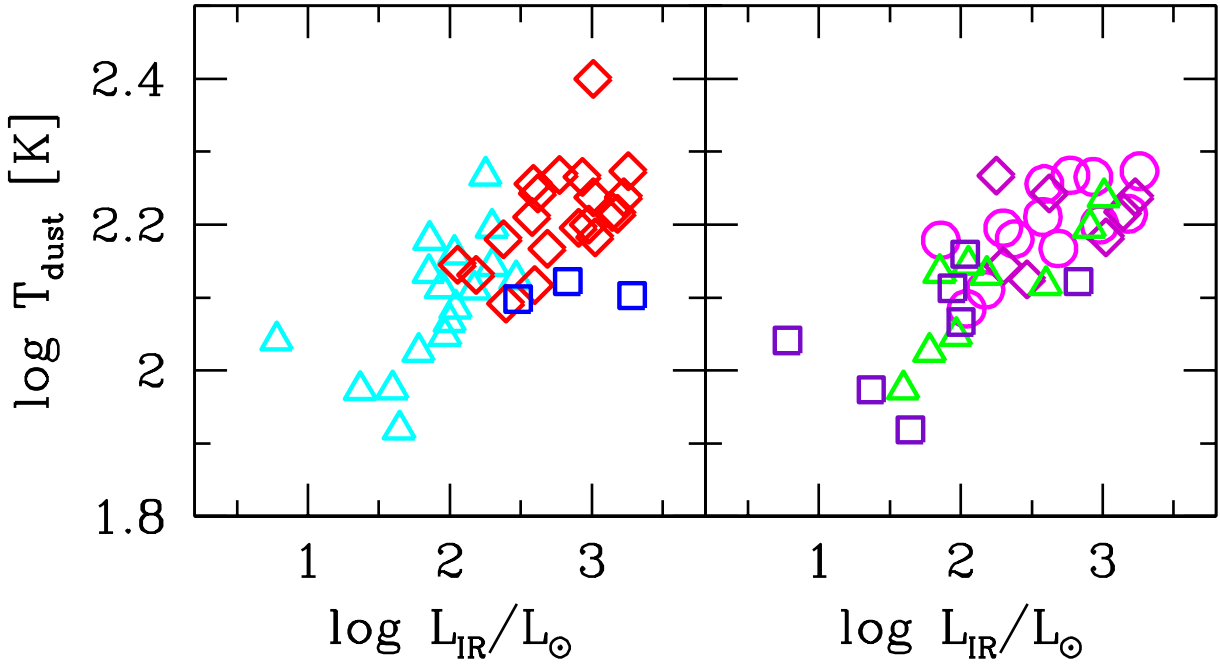}
\caption{Dust temperature versus IR luminosity for dust type (left) and morphology (right).
Symbols as in Fig. 9.}
\end{figure}

\begin{figure}
\figurenum{13}
\includegraphics[bb = 100 100 600 600, angle=0]{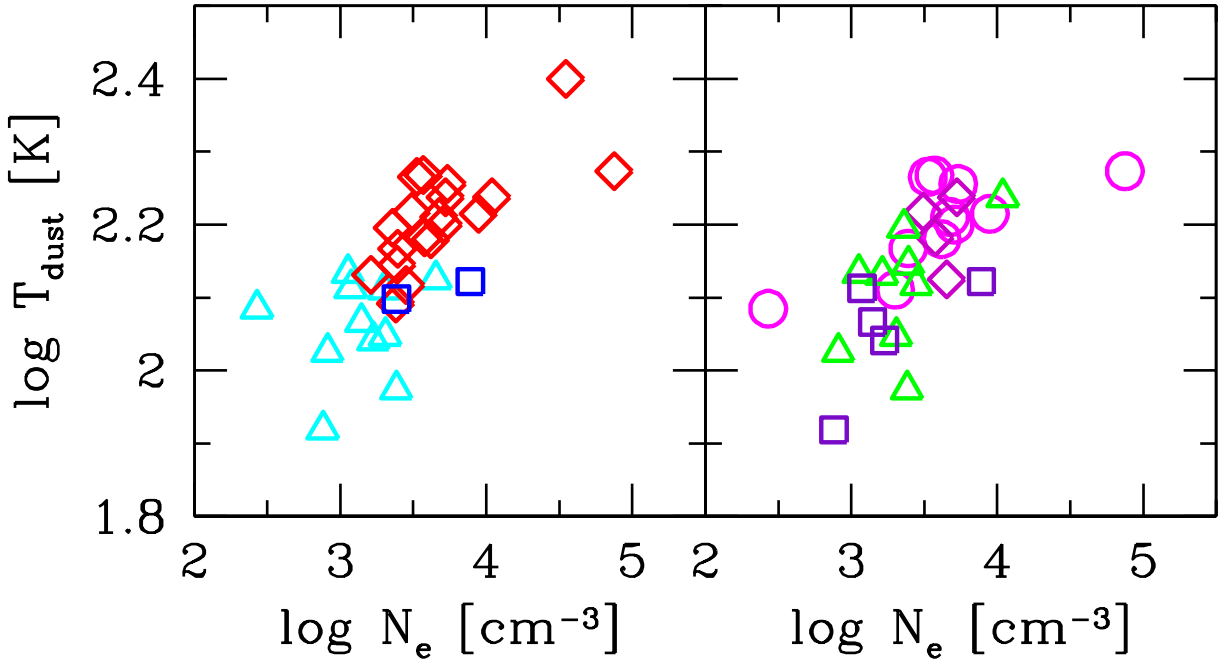}
\caption{Dust temperature versus electron density for dust type (left) 
and morphology (right). Symbols as in Fig.~9.}
\end{figure}

\begin{figure}
\figurenum{14}
\includegraphics[bb = 100 100 600 600, angle=0]{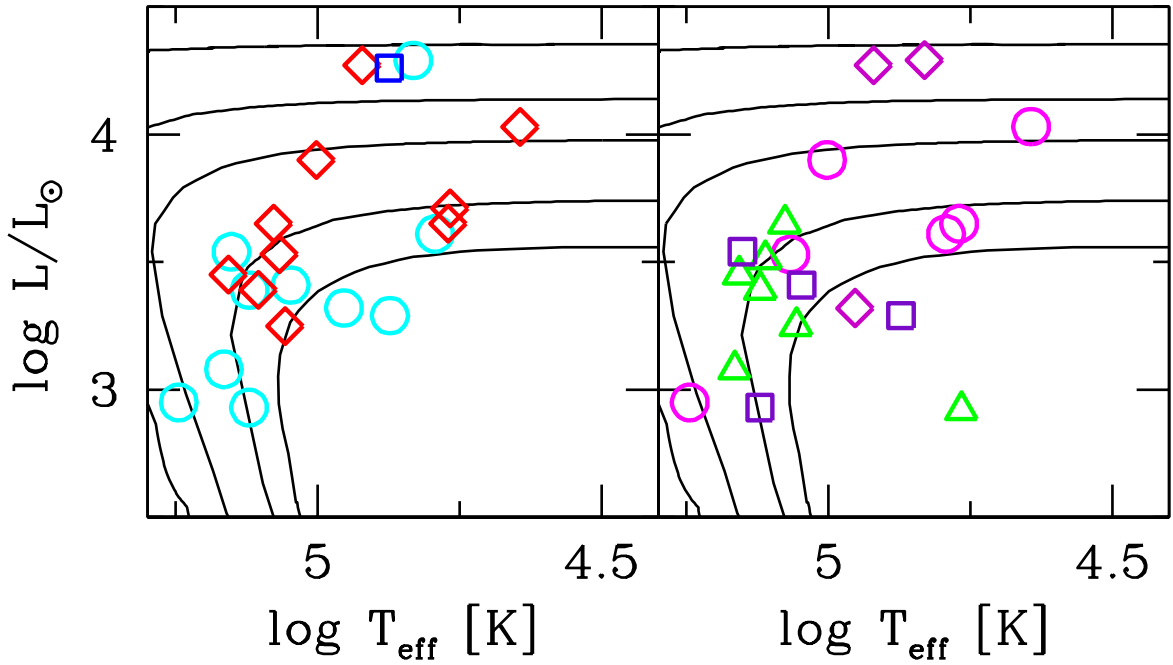}
\caption{The loci of the central stars of a subsample of the studied PNe on the HR plane. Symbols of the observed PNe
are as in Fig.~9.}
\end{figure}

\begin{figure}
\figurenum{15}
\vspace*{-15mm}
\plotone{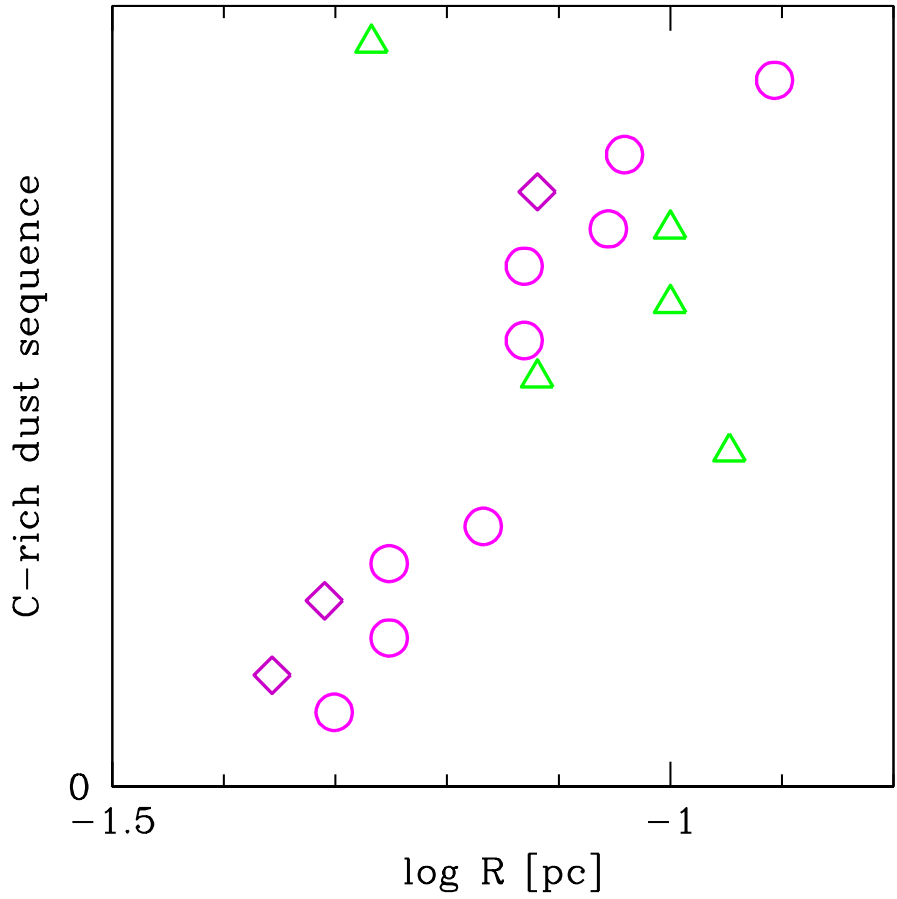}
\caption{Schematic evolutionary sequence of CRD PNe, where we plotted the order of apparent 
evolution of the carbonaceous dust features, as in Figure
7a, b, and c, versus the photometric radii of the nebulae. Circles are 
round PNe, diamonds are elliptical PNe, and triangles are bipolar core PNe.}
\end{figure}

\end{document}